\documentclass[reprint, superscriptaddress, amsmath, amssymb, aps, prb]{revtex4-2}

\usepackage{amsmath}
\usepackage{amssymb}
\usepackage[colorlinks=true, allcolors=blue]{hyperref}
\usepackage{indentfirst}
\usepackage{commath}
\usepackage{graphicx}
\usepackage[caption=false,position=bottom,labelfont={bf}]{subfig}
\usepackage{overpic}
\usepackage{dcolumn}
\usepackage{bm}

\usepackage{xcolor}
\definecolor{light-red}{rgb}{0.8,0.15,0.5}
\definecolor{dark-green}{rgb}{0.275,0.47,0.47}
\definecolor{medium-blue}{rgb}{0,0,1}
\hypersetup{
colorlinks, linkcolor={medium-blue},
citecolor={medium-blue}, urlcolor={medium-blue}
}

\begin{document}

\title{Phase fluctuations in two-dimensional superconductors and pseudogap phenomenon}

\author{Xu-Cheng Wang}
\affiliation{State Key Laboratory of Surface Physics, Fudan University, Shanghai 200433, China}
\affiliation{Center for Field Theory and Particle Physics, Department of Physics, Fudan University, Shanghai 200433, China}

\author{Yang Qi}
\email{qiyang@fudan.edu.cn}
\affiliation{State Key Laboratory of Surface Physics, Fudan University, Shanghai 200433, China}
\affiliation{Center for Field Theory and Particle Physics, Department of Physics, Fudan University, Shanghai 200433, China}
\affiliation{Collaborative Innovation Center of Advanced Microstructures, Nanjing 210093, China}

\date{\today}

\begin{abstract}
  We study the phase fluctuations in the normal state of generic two-dimensional superconducting systems with $s$-wave pairing.
  The effect of phase fluctuations of the pairing fields can be dealt with perturbatively using disorder averaging,
  after we treat the local superconducting order parameter as a static disordered background.
  It is then confirmed that the phase fluctuations above the two-dimensional Berezinskii-Kosterlitz-Thouless transition lead to a significant broadening of the single-particle spectrum, giving birth to the pseudogap phenomenon.
  Quantitatively, the broadening of spectral weights near the BCS gap is characterized by the ratio of the superconducting coherence length
  and the spatial correlation length of the superconducting pairing order parameter.
  Our results are tested on the fermionic attractive-$U$ Hubbard model on the square lattice,
  using the unbiased determinant quantum Monte Carlo method and stochastic analytic continuation.
\end{abstract}

\maketitle

\section{Introduction}

Over the past few decades,
the phemomenon of high-$T_c$ superconductivity has remained
a mysterious but intriguing problem in the field of condensed-matter physics~\cite{Cuprate1986,
TsueiRMP2000,LNW-HighTc-RMP2006,Keimer2015Nature,Proust2018Review,Arovas2022Review}.
It has long been experimentally confirmed that,
in the phase diagram of underdoped cuprate superconductors,
there exists a pseudogap regime located above the transition temperature $T_c$
and below a characteristic temperature $T^\ast$,
where an energy gap evolves smoothly from the superconducting gap at $T_c$~\cite{Timusk1999,
Vedeneev2021,sobota_angle-resolved_2021}.
The onset of pseudogap behavior below $T^\ast$ remains a controversial issue:
Many scenarios including other order parameters competing with superconductivity,
e.g., stripe order or antiferromagnetic order, have been put forward
from both theoretical~\cite{chakravarty_hidden_2001, GuUPG2005, varma_theory_2006,
grilli_fermi_2009}
and empirical~\cite{li_unusual_2008, he_single-band_2011, chang_direct_2012}
perspectives.
In some scenarios, the pseudogap phase is regarded as a symmetry-breaking phase,
and the temperature $T^\ast$ as a phase transition temperature~\cite{chakravarty_hidden_2001, varma_theory_2006, li_unusual_2008}.
The physical origin of the pseudogap has been a controversial topic till now.

On the other hand, it is now broadly acknowledged that up to a characteristic temperature of $T_\mathrm{Ong}<T^\ast$~\cite{franz_importance_2007, hashimoto_energy_2014},
which can be as high as 140 K in certain materials,
the strong phase fluctuations of the superconducting order parameter
are responsible for the pseudogaplike behavior.
Evidences of fluctuating superconductivity are observed in photoemission measurements on cuprate $\text{Bi}_2\text{Sr}_2\text{Ca}\text{Cu}_2\text{O}_{8+\delta}$ (Bi2212) over a temperature range of $\Delta T\sim\left(0.3-0.5\right)T_c$ near both the node and the antinode~\cite{hashimoto_direct_2015,kondo_point_2015,chen_incoherent_2019}.
It is expected that the phase fluctuation in cuprate superconductors is partially enhanced by the effective two-dimensionality of the materials. 
In particular, in an ultrathin superconducting film where the couplings between layers
are comparably weak, the superconducting transition falls into
the Berezinskii-Kosterlitz-Thouless (BKT) universality class~\cite{kosterlitz_ordering_1973, kosterlitz_critical_1974},
which is driven by phase fluctuation or equivalently by proliferation of vortices.
In Refs.~\cite{xu_vortex-like_2000, balci_nernst_2003, wang_nernst_2006},
vortexlike excitations were directly observed in the pseudogap regime
by measuring the Nernst effect of moving vortices,
which destroyed the phase coherence in the normal phase
while conserving the superconducting gap.
These experiments provide compelling evidence for
the scenario of phase fluctuation below $T_\mathrm{Ong}$.
\citet{bergeal_pairing_2008} measured the Josephson effect of
fluctuating pairs in a trilayer junction,
and they concluded that the pairing fluctuations
only survived in a restricted range of temperature above $T_c$
if the samples were quite clean,
indicating that the upper boundary of the phase-fluctuating regime $T_\mathrm{Ong}$
might be related to the disordering of the system.

A mass of theories have been developed to describe
such kinds of phase-disordered superconductors~\cite{kwon_effect_1999, franz_phase_1998, kwon_observability_2001,
franz_algebraic_2001, curty_thermodynamics_2003}:
\citet{kwon_effect_1999} proposed
a effective low-energy theory of fermionic quasiparticles coupled to
the phase fluctuations of the superconducting order parameter,
and it is shown that
the vortex-pair unbounding near $T_c$ will produce a pseudogaplike feature.
\citet{franz_phase_1998} coupled the $d$-wave quasiparticles
to the fluctuating supercurrents due to unbound vortex-antivortex pairs,
and pointed out that only the transverse phase fluctuations are important
in determining the spectral properties
while longitudinal fluctuations are unimportant at all temperatures.
\citet{curty_thermodynamics_2003} took both amplitude and phase fluctuations
of the pairing fields into account
and properly explained the specific heat and magnetic susceptibility experiments.
So far, the theoretical progress had mainly focused on the phase fluctuation of $d$-wave superconducting order parameters
and its interplay with the nodal quasiparticles,
which are specific to cuprate superconductors.
However, the existence of superconducting fluctuations and pseudogap behavior is expected to be a generic feature of two-dimensional (2D) superconductors,
because of the BKT nature of the 2D superconducting phase transitions.
Correspondingly, a generic and simple theoretical framework describing these phenomena is still lacking.

In this paper, we provide an intuitive way to derive 2D pseudogap phenomenon,
treating thermal fluctuations and static disorders of the phase of pairing order parameters on an equal footing.
Assuming that the phase of the superconducting order parameter 
fluctuates only classically
because only classical fluctuations play an important role at the finite temperatures $T_c<T<T_{\text{Ong}}$,
we calculate its correction to the electronic self-energy
using the standard disorder-averaging method.
To the lowest order of perturbative expansion,
we obtain the analytic form of the single-particle spectral functions.
Consistent with previous works~\cite{kwon_effect_1999},
we find that
the broadening of the single-particle spectrum is determined by
the ratio of two characteristic lengths in the system:
the superconducting coherence length $\xi_\mathrm{BCS}$ in Ginzburg-Landau theory
and the correlation length $\xi$ of the disordered phases.
Furthermore, we demonstrate this result numerically in an attractive-$U$ Hubbard model, 
which is known to realize a superconducting phase at finite doping,
using the unbiased determinant quantum Monte Carlo (DQMC) method.
Our numerical simulation confirms these theoretical results.

\section{Model and Methods}

In this section, we propose an effective Hamiltonian with the fermionic degrees
of freedom coupled to fluctuating superconductivity
as in Eqs.~(\ref{eq:hamiltonian_free}) and~(\ref{eq:hamiltonian_interaction}):
\begin{equation} \label{eq:hamiltonian_free}
      \hat{H}_0 = \sum_\sigma \int\mathrm{d^2}r\
      \hat\psi^\dagger_\sigma(r) \left(-\frac{\bigtriangledown^2}{2m}-\mu\right) \hat\psi_\sigma(r),
\end{equation}
\begin{equation} \label{eq:hamiltonian_interaction}
      \hat{V} = \int\mathrm{d^2}r\ \left[\Delta(r)\hat\psi^\dagger_\uparrow(r) \hat\psi^\dagger_\downarrow(r) + \text{h.c.}\right],
\end{equation}
where $\hat{H}_0$ describes the free spinful fermion in two dimensions
with $\sigma$ labeling the spin degrees of freedom,
and $\hat{V}$ represents the coupling between electrons and superconducting fluctuations $\Delta(r)$.
$\Delta(r)$ are complex fields of the superconducting order parameter,
and we assume that the pairing is $s$-wave-like to simplify the problem.
The model above is a generic phenomenological description to an $s$-wave superconductor,
and we have put the microscopic mechanism of superconducting pairing aside.
The focus of our research is on how the disappearance of phase coherence
above the transition temperature $T_c$ affects the quasiparticle spectrum.

Here, we only consider the classical fluctuations of $\Delta$, and we assume $\Delta(r)$ is independent of time.
This is because in the vicinity of the BKT transition temperature,
only the classical fluctuations are important in the low-energy effective theory,
as the correlation length $\xi$ diverges while the correlation in the imaginary-time direction is cut off by the inverse temperature $\beta=(k_BT)^{-1}$.
Therefore, in a low-energy effective theory with a cutoff length scale $a$ satisfying $v_F\beta<a<\xi$, all temporal fluctuations of $\Delta(r)$ have been integrated out.

In momentum space, we can rewrite the Hamiltonian as
\begin{equation} \label{eq:hamiltonian}
      \begin{aligned}
            \hat{H} &= \sum_\sigma \int \frac{\mathrm{d^2}k}{\left(2\pi\right)^2}\
                       \xi_k \hat\psi^\dagger_\sigma(k) \hat\psi_\sigma(k) \\
                    &+ \iint \frac{\mathrm{d^2}k}{\left(2\pi\right)^2}
                       \frac{\mathrm{d^2}k^\prime}{\left(2\pi\right)^2}\
                       \left[\Delta_{kk^\prime} \hat\psi^\dagger_\uparrow(k)
                       \hat\psi^\dagger_\downarrow(k^\prime)+\text{h.c.}\right],
      \end{aligned}
\end{equation}
where $\xi_k=k^2/2m-\mu$ is the dispersion relation of electrons,
$\mu=k_F^2/2m$ is the chemical potential, and $k_F$ is the Fermi momentum as usual.
The Fourier form of the pairing fields reads $\Delta_{kk^\prime} =
\int\mathrm{d^2}r \Delta(r) e^{-i(k+k^\prime)r}$.
It is notable that the space dependency of $\Delta(r)$ breaks the translational
invariance, so the momentum is not conserved during the scattering processes induced by $\Delta_{kk^\prime}$.

In order to take into account the phase fluctuations of the pairing,
we separate the amplitude and phase components of the pairing order parameter as $\Delta(r) = |\Delta(r)|e^{i\theta(r)}$.
It is assumed that near the transition temperature of superconductivity,
only the phase component fluctuates significantly and the amplitude part possesses a well-defined value
over the whole space $|\Delta(r)|=\Delta_0$.
As a consequence, the superconducting pairing fields can be approximated as some kind of classical disordered backgrounds,
and we can resort to the technique of disorder average~\cite{mahan_many_2000},
which was originally developed to deal with impurity scattering problem, combined with perturbation theory,
to include the contribution of phase fluctuations order by order.

The disorder-average conditions are as follows
\begin{equation} \label{eq:1-order-average}
      \overline{\Delta(r)} = 0,
\end{equation}
\begin{equation} \label{eq:2-order-average}
      \overline{\Delta(r) \Delta^*(r^\prime)}
      = {\Delta_0}^2\ e^{-\frac{|r-r^\prime|^2}{2\xi^2}},
\end{equation}
where the overline represents the average
over different configurations of $\Delta(r)$.
$\xi$ characterizes the spatial correlation length of the pairing.

In Eq.~(\ref{eq:2-order-average}) we have imposed a Gaussian-like correlation function
which is not compulsory and just a choice for the convenience of calculation.
Indeed, by numerically carrying out the detailed calculations in the next section, we find the specific line shape of correlations unrelative to
the long wave behavior of the theory if we are satisfied with
the physics in the vicinity of $T_c$ where $k_F\xi \gg 1$.
This is, in turn, an important advantage of the disorder-averaging approach in that
it depends little on the microscopic dynamics of superconducting fields
and thus has the potential to treat the thermal fluctuations and static disorders in the material on the same footing.
\section{Theoretical Analysis}
\label{sec:theory}
We base our theoretical analysis on the perturbative expansion of the pairing amplitude $\Delta_0$.
This is justified if $\Delta_0$ is much smaller than the Fermi energy (or the bandwidth).
To the lowest order of nonvanishing expansion, which corresponds to a second-order scattering process,
we write down the self-energy of electrons by considering the pairing fields as some scattering backgrounds
\begin{equation} \label{eq:perturbative-expansion}
    \Sigma(k, k^\prime, \omega) = \int \frac{\mathrm{d^2}p}{(2\pi)^2} \frac{\Delta_{kp}\Delta^*_{pk^\prime}}{\omega+\xi_p+i\epsilon},
\end{equation}
where $\epsilon$ is a positive infinitesimal quantity.
Note that the propagator has a pole at $\omega = -\xi_k$,
indicating the number of fermion particles is not conserved at the interaction vertex.

The disorder averaging is then straightforward to perform
by simply replacing the correlation function of pairing fields with the averaged one
using the Fourier form of Eq.~(\ref{eq:2-order-average}).
Furthermore, it can be easily verified that, after the averaging operation, the translation symmetry is recovered
as long as the correlation of pairing fields depends only on the displacement of two space points.
We obtain
$\overline{\Sigma(k,k',\omega)}=\Sigma(k,\omega)\delta_{k,k'}$ and
\begin{equation} \label{eq:self-energy}
      \Sigma(k, \omega)
            = 2\pi\xi^2{\Delta_0}^2 \int \frac{\mathrm{d^2}p}{(2\pi)^2}
              \frac{e^{-\frac{1}{2}\xi^2(k+p)^2}}{\omega+\xi_p+i\epsilon}.
\end{equation}

In the limit of infinite correlation length $\xi\to\infty$, e.g., deep in the superconducting phase,
the self-energy in Eq.~(\ref{eq:self-energy}) degenerates to the mean-field result in the BCS theory as expected.
While for large but finite correlation length, which corresponds to the normal state where $T>T_c$
and the phase fluctuations are un-neglectable,
the paired momentum will fluctuate near $p=-k$ and get truncated by roughly $\xi^{-1}$.
This is the main feature beyond the usual BCS theory by including the contributions of phase fluctuations.
\begin{widetext}
      \begin{align}
            \mathrm{Im}\ \Sigma(k, \omega)
            &= -2\pi^2\xi^2{\Delta_0}^2 \int \frac{\mathrm{d^2}p}{(2\pi)^2} e^{-\frac{1}{2}\xi^2(k+p)^2} \delta(\omega+\xi_p) \label{eq:self-energy-imag} \\
            &= -\pi m\xi^2{\Delta_0}^2\cdot e^{-\xi^2k^2+m\xi^2(\omega+\xi_k)} \times I_0\left(\xi^2k^2\sqrt{1-\frac{2m}{k^2}(\omega+\xi_k)}\right), \quad \text{with}\ \omega<\mu. \label{eq:self-energy-imag-solution}
      \end{align}
\end{widetext}

In principle, Eq.~(\ref{eq:self-energy}) is satisfactory
for the numerical determination of the single-particle properties of electrons.
However, in attempt to make the expression more concise,
we can take a closer look at the imaginary part of the self-energy,
as shown in Eqs.~(\ref{eq:self-energy-imag}) and~(\ref{eq:self-energy-imag-solution}),
where we have performed the integral over the 2D momentum exactly
and $I_0(z)$ represents the zero-order modified Bessel function of the first kind.
The well-defined imaginary component of self-energy
indicates a finite lifetime of BCS quasiparticles in the presence of phase fluctuations,
which will contribute to the significant broadening of spectral function in the pseudogap regime, as we discuss below.
Notice that we are interested in the upper neighborhood of $T_c$, and
Eq.~(\ref{eq:self-energy-imag-solution}) can be further simplified
if we take the asymptotic expression of the Bessel function in the case of $k_F\xi \gg 1$, and this leads to
\begin{equation} \label{eq:self-energy-normal}
      \mathrm{Im}\ \Sigma(k, \omega)
            = -\pi{\Delta_0}^2 \frac{m\xi}{(2\pi)^{1/2}k} \
              e^{-\frac{m^2\xi^2}{2k^2}(\omega+\xi_k)^2}.
\end{equation}
It can be seen that the imaginary part of the self-energy obeys a Gaussian distribution with the standard deviation $\frac{k}{m\xi}$.
This motivates us to rewrite the self-energy as
\begin{equation} \label{eq:self-energy-approx}
      \Sigma(k, \omega) = \frac{{\Delta_0}^2}{\omega+\xi_k+i\frac{k}{m\xi}}.
\end{equation}
To obtain Eq.~(\ref{eq:self-energy-approx}),
we first substitute the normal distribution in Eq.~(\ref{eq:self-energy-normal}) with a Lorentzian,
since we assume that the specific form of correlations in Eq.~(\ref{eq:2-order-average})
should not affect the low-lying physics for a generic superconductor.
The real part of the self-energy is then determined from the Kramers-Kronig relation.
It has also been examined numerically that, when $k_F\xi \gg 1$ is satisfied,
the approximated self-energy Eq.~(\ref{eq:self-energy-approx}) reproduces similar real and imaginary components as given by Eq.~(\ref{eq:self-energy}).
From Eq.~(\ref{eq:self-energy-approx}), we can clearly identify the interplay between electrons and fluctuating superconductivity.
To the lowest order of nonvanishing corrections,
the phase incoherence in the normal state of superconductors offers a nonzero scattering rate of $k/m\xi$ to the BCS quasiparticles.
We demonstrate below that this scattering process gradually smears the BCS quasiparticle peaks and generally leads to the emergence of the pseudogap.

\subsection{Pseudogap behavior}
Using Eq.~(\ref{eq:self-energy-approx}), 
we are ready to compute the retarded Green's function $G(k,\omega)$ in the presence of phase fluctuations:
\begin{equation} \label{eq:greens-function}
      \begin{aligned}
            G(k,\omega) &= \left[ \omega - \xi_k - \Sigma(k,\omega) \right]^{-1} \\
                        &= \frac{ \omega + \xi_k+ 2i\gamma_k }{ \omega^2 - E_k^2 + 2i\gamma_k \left(\omega-\xi_k\right) },
      \end{aligned}
\end{equation}
where we define $\gamma_k=\frac{k}{2m\xi}$, 
and $E_k = \sqrt{\xi_k^2+{\Delta_0}^2}$ is the electron's dispersion relation in the BCS theory.
Equation~(\ref{eq:greens-function}) can be viewed as the Green's function of 
finite-lifetime quasiparticles described by a non-Hermitian one-body Hamiltonian~\cite{nagai_dmft_2020}.
In this case, the spectrum is embedded in the pole of the Green's function, which is, in general, complex.
The real part of the complex pole determines the quasiparticle's dispersion,
while the imaginary part leads to a finite lifetime of quasiparticles. 

\begin{figure}[htbp]
      \subfloat{
            \label{fig:analytic-greens-function}
            \begin{overpic}[width=0.9\columnwidth]{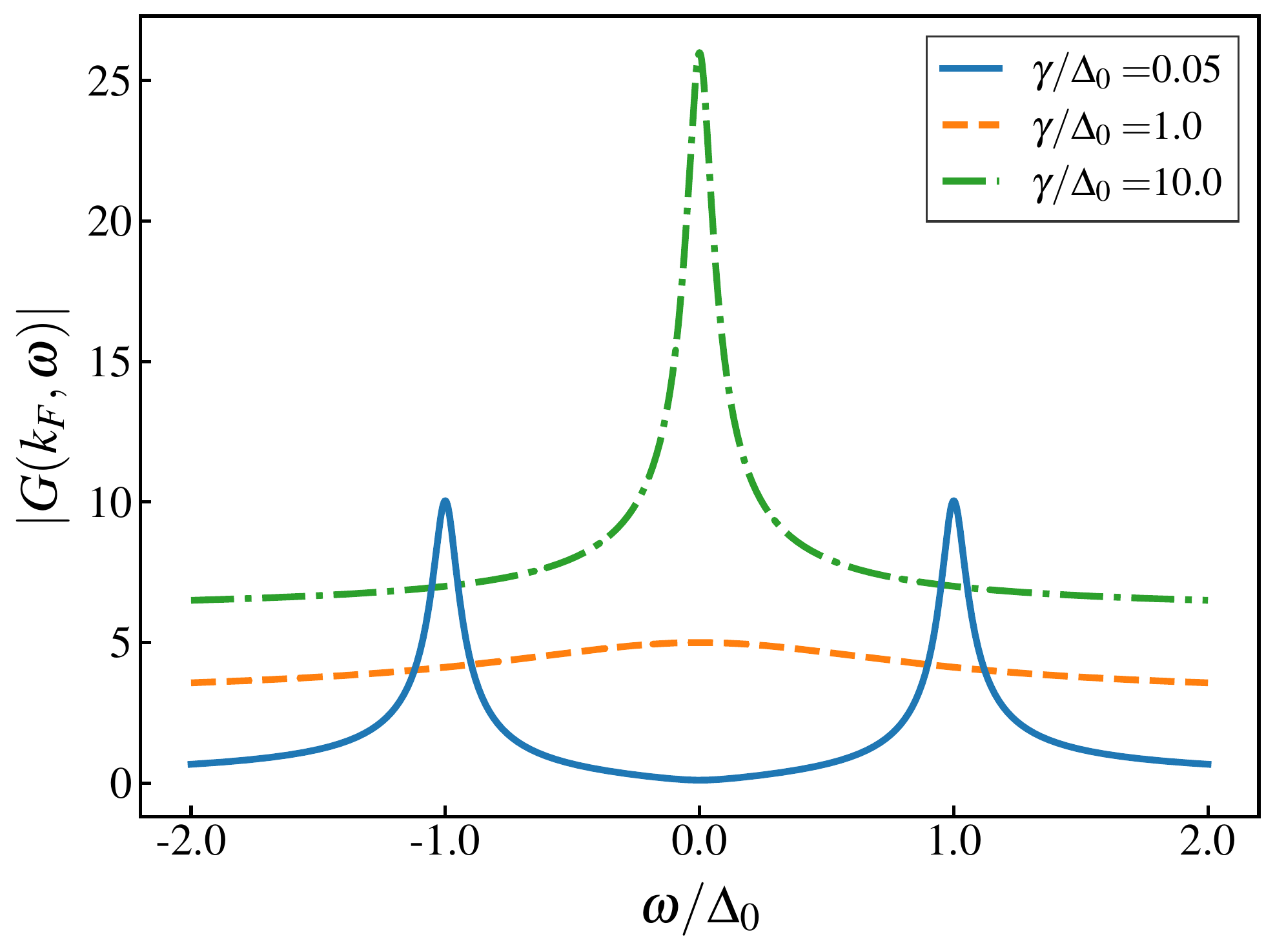}
                  \put(15,65){\large(a)}
            \end{overpic}
      }\\
      \subfloat{
            \label{fig:analytic-dos}
            \begin{overpic}[width=0.9\columnwidth]{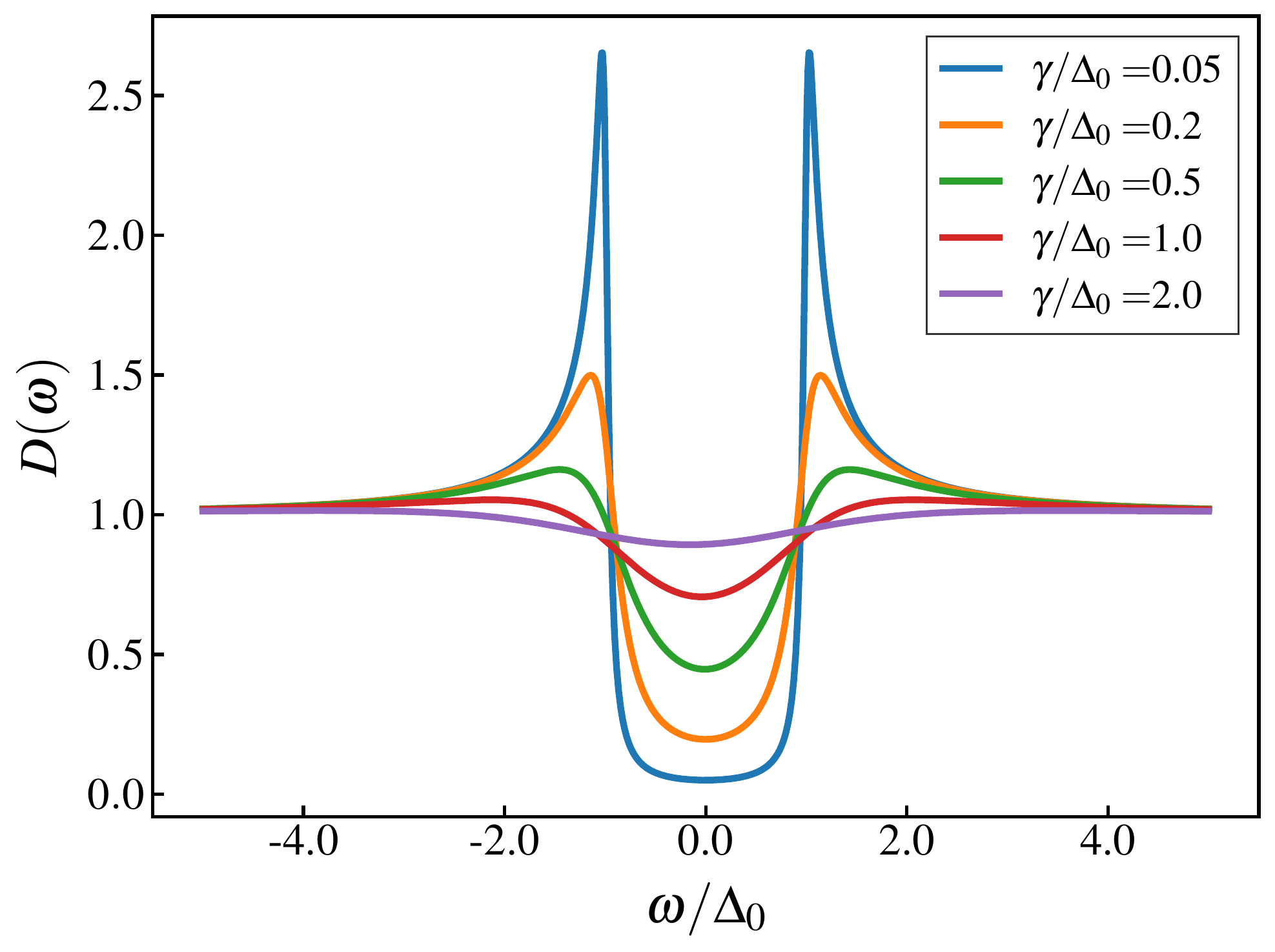}
                  \put(15,65){\large(b)}
            \end{overpic}
      }
      \caption{
            \label{fig:analytic-results}
            (a) Energy dependence of the amplitude of the retarded Green's function $|G(k_F,\omega)|$ at the Fermi surface
            for $\gamma/\Delta_0=$ 0.05, 1.0, and 10.0.
            (b) Density of states $D(\omega)$ for $\gamma/\Delta_0=$ 0.05, 0.2, 0.5, 1.0, and 2.0.
            It is clear that the gap is gradually filled as the phase coherence is destroyed by thermal fluctuations.
      }
\end{figure}

This motivates us to find the complex pole in Eq.~(\ref{eq:greens-function}) by solving the following equation:
\begin{equation}
      \omega^2 - E_k^2 + 2i\gamma_k \left(\omega-\xi_k\right) = 0,
\end{equation}
which will generally give two solutions:
\begin{equation} \label{eq:complex-pole-solution}
      \omega = -i\gamma_k \pm \sqrt{E_k^2-\gamma_k^2+2i\gamma_k\xi_k}.
\end{equation}
If we consider the quasiparticles near the Fermi surface where $\xi_k=0$,
then Eq.~(\ref{eq:complex-pole-solution}) turns into
\begin{equation} \label{eq:complex-pole-solution-fs}
      \omega = -i\gamma \pm \sqrt{{\Delta_0}^2-\gamma^2}
\end{equation}
and $\gamma=\frac{v_F}{2\xi}$ is proportional to the inverse of the correlation length.
For $\gamma/\Delta_0\ll1$, i.e., the phase-coherent region near the critical temperature,
the Green's function has two complex poles with opposite real components $\pm\Delta_0$,
which verifies the formation of BCS quasiparticles, and the spectrum should be fully gapped.
While for $\gamma/\Delta_0\gg1$, Eq.~(\ref{eq:complex-pole-solution-fs}) is purely imaginary
and the real component is zero, which is exactly the case of the Fermi liquid.
In the intermediate region $\gamma/\Delta_0\sim1$ where the phase fluctuations are relevant,
a crossover from Fermi liquid to the BCS physics, e.g., the pseudogap, should occur.
The plot of the retarded Green's function for typical values of $\gamma/\Delta_0$
is shown in Fig.~\ref{fig:analytic-greens-function}.

The pseudogap phenomenon will naturally emerge if we compute the fermionic spectral function $A(k,\omega)$:
\begin{equation} \label{eq:spectral-function}
      \begin{aligned}
            A(k,\omega) &= -2\ \mathrm{Im}G(k,\omega) \\
                        &= \frac{4{\Delta_0}^2\gamma_k}
                        {\left(\omega^2-E_k^2\right)^2+4\gamma_k^2\left(\omega-\xi_k\right)^2}
      \end{aligned}
\end{equation}
and the density of states (DOS) $D(\omega) = \int \frac{\mathrm{d}^2k}{(2\pi)^2} A(k,\omega)$.
Fig.~\ref{fig:analytic-dos} illustrates the energy dependence of $D(\omega)$
in the pseudogap regime with $\gamma/\Delta_0\lesssim1$.
We can see explicitly the gap-filling behavior as phase fluctuations are enhanced.

\subsection{Broadening of spectral peaks}
Following our analysis from Eq.~(\ref{eq:greens-function}) to Eq.~(\ref{eq:spectral-function}),
it becomes obvious that when phase fluctuations are suppressed in the vicinity of critical temperature and $\gamma/\Delta_0\ll1$,
the spectral function $A(k=k_F,\omega)$ peaks at $\omega=\pm\Delta_0$ with a finite half-width $\Delta\omega$.
More specifically, from Eq.~(\ref{eq:spectral-function}) we can estimate
\begin{equation} \label{eq:half-width-relation}
      \Delta\omega = \gamma + O(\gamma)
                   = \frac{\pi}{2} \Delta_0 \frac{\xi_{\mathrm{BCS}}}{\xi} + O(\xi^{-1}),
\end{equation}
where $\xi_{\mathrm{BCS}} = \frac{v_F}{\pi\Delta_0}$ is the superconducting coherence length~\cite{annett_superconductivity_2003},
which characterizes the physical size of the Cooper pair in the BCS theory, and $v_F$ is the Fermi velocity.
The contribution of phase incoherence to the broadening of spectral peaks
is thus characterized by the ratio between
the correlation length $\xi$ of the pairing order parameter and the superconducting coherence length $\xi_{\mathrm{BCS}}$.
According to our previous assumptions, when the temperature is in the pseudogap regime $T_c<T<T_{\text{Ong}}$,
the superconducting gap $\Delta_0$ is independent of temperature,
while $\gamma$ is highly sensitive to the changes in temperature through $\xi$;
e.g., for 2D BKT transition, the correlation length decreases exponentially above $T_c$ [ see the scaling relation in Eq.~\eqref{eq:scaling} ].
As the temperature increases, i.e., increase $\gamma$, the half-width in Eq.~(\ref{eq:half-width-relation}) grows up,
and the peaks are smeared gradually. When $\gamma/\Delta_0\gtrsim1$,
the two peaks merge into each other and eventually the gap in the single-particle spectrum is closed.
Following a straightforward calculation, we find that $A(k=k_F,\omega)$ reaches its maximum at $\omega=\pm\sqrt{{\Delta_0}^2-2\gamma^2}$,
and the pseudogap closes when $\gamma/\Delta_0=\frac{1}{\sqrt{2}}$, which is less than 1 slightly due to the finite broadening of BCS peaks.

Therefore, Eq.~(\ref{eq:half-width-relation}) can serve as a quantitative indicator
to measure the effect of phase fluctuations on the spectral properties of electrons in such phase-disordered superconductors.
The half-width of spectral peaks $\Delta\omega$ and the correlation length $\xi$ can also be directly measured by numerical simulations,
which makes it possible to confirm our theory on generic lattice models.

Furthermore, we expect a similar analysis can also be applied to $d$-wave superconductors.
For a system with $d_{x^2-y^2}$ pairing, the superconducting gap $\Delta_0(k)$ becomes momentum dependent,
and the ratio $\gamma/\Delta_0$ may vary along the Fermi surface.
Consequently, in the pseudogap state of $d$-wave superconductors,
the phase fluctuations may open a pseudogap on part of the Fermi surface where the ratio is small, while the gap remains closed on other part of the Fermi surface where the ratio is large,
and hence leads to Fermi arcs~\cite{nagai_dmft_2020}.
We argue that, in contrast with the pseudogap emerging in the underdoped regime and near antinodes of cuprates,
the pseudogap induced by fluctuating superconductivity, accompanied by the Fermi arcs, is, however, general and ubiquitous in 2D superconductors with $d$-wave pairing.
This is supported by not only certain photoemission measurements of Bi2212~\cite{hashimoto_direct_2015,kondo_point_2015}, which indicate that the fluctuation-induced pseudogap exists above the entire superconducting dome,
but also numerical studies of 2D superconductors with pure superconducting pairing~\cite{singh_fermi_2022}.
We leave a detailed study of this possibility to future works.

It is also notable that although we have based our calculations on the theory of free fermion with a circular Fermi surface,
the relation in Eq.~(\ref{eq:half-width-relation}) can be quite general
as it should be independent of the shape of the Fermi surface or the details of the microscopic theory of superconductivity,
e.g., a specific form of the decay of correlation functions.
Hence, it is expected to correctly capture the low-energy properties of
the generic $s$-wave superconductors in the pseudogap regime even when there exists strong interaction among electrons, as we see numerically in the next section.

\section{DQMC Simulations}

In order to verify the role of phase fluctuations and the relation in Eq.~(\ref{eq:half-width-relation}),
we study the fermionic Hubbard model with attractive on-site interaction on the square lattice
via the DQMC algorithm.
The model Hamiltonian is
\begin{equation} \label{eq:hubbard-model}
      \begin{aligned}
            H = &-t \sum_{\left<i,j\right>, \sigma}
                  \left(c_{i\sigma}^\dagger c_{j\sigma}
                  +c_{j\sigma}^\dagger c_{i\sigma}\right)
                  + \mu \sum_{i\sigma} n_{i\sigma} \\
                &- \abs{U} \sum_{i} \left(n_{i\uparrow}-\frac{1}{2}\right)
                  \left(n_{i\downarrow}-\frac{1}{2}\right),
      \end{aligned}
\end{equation}
where $t$ describes the hopping amplitude between nearest-neighbor sites,
$|U|$ is the strength of attractive on-site interaction, and $\mu$ is the chemical potential.
It is proved that this model is free of the notorious sign problem for any value of doping and interaction strength.
This is guaranteed by the time-reversal symmetry of the model action after the Hubbard-Stratonovich transformation~\cite{wu_sufficient_2005}.

From the historical view, the attractive-$U$ Hubbard model has aroused considerable interest
because it shows a BKT transition to an $s$-wave superconducting state.
The phase diagram, especially the location of the superconducting transition,
was systematically studied in Refs.~\cite{moreo_two-dimensional_1991, paiva_critical_2004}.
At half-filling, i.e., $\mu=0$, the transition temperature is suppressed to zero
due to the degeneracy of the charge density order and the superconducting one.
However, away from half-filling, the charge fluctuations are suppressed,
leading to a BKT superconducting transition with a finite transition temperature $T_c$~\cite{paiva_critical_2004, vilk_attractive_1998}.
Furthermore, signals of performed pairs have been found in the normal state within a certain range of parameters through the measurements of spin susceptibility,
which is related to the pseudogap behavior in high-$T_c$ superconductors~\cite{randeria_pairing_1992, trivedi_deviations_1995}.

In this section, we confirm the existence of the pseudogpap phase in the attractive-$U$ Hubbard model through large-scale DQMC simulations,
and the gap-filling behavior is directly observed in both spectral functions and DOS.
In addition, we attempt to verify that the phase fluctuations contribute to the broadening of fermionic spectral weights as predicted in Eq.~(\ref{eq:half-width-relation}).
To see this, we first determine the BKT transition temperature $T_c$ of the superconductivity and obtain the temperature dependence of the correlation length $\xi$ by scaling analysis.
Then, the half-width of the spectral weights is directly read from the fermionic spectral functions,
which can be recovered from imaginary-time correlation functions using stochastic analytic continuation (SAC)~\cite{sandvik_stochastic_1998,beach_identifying_2004,shao_nearly_2017,shao_progress_2023}.
We notice that the pseudogap phenomenon has been observed using this method in models with fermionic quantum critical points~\cite{WLJiang2022,YCHao2022}.

\subsection{Determination of $T_c$ and data collapse}
In order to determine the transition temperature $T_c$, we define the $s$-wave pairing correlation function $P_s$~\cite{moreo_two-dimensional_1991} as
\begin{equation}
      P_s = \left<\Delta^\dagger\Delta+\Delta\Delta^\dagger\right>,
\end{equation}
where $\Delta^\dagger = \frac{1}{\sqrt{N}}\sum_ic^\dagger_{i\uparrow}c^\dagger_{i\downarrow}$.

Near the critical point of the BKT transition, the correlation length $\xi$ and the pairing $P_s$ should behave as~\cite{kosterlitz_critical_1974}
\begin{equation} \label{eq:scaling}
      \xi \propto \mathrm{exp}\left(at_r^{-1/2}\right)
      \quad \text{and} \quad P_s \propto \xi^{2-\eta},
\end{equation}
where $a$ is a nonuniversal constant and $t_r=\left(T-T_c\right)/T_c$ is the reduced temperature.
For the BKT transition, the critical exponent $\eta\left(T_c\right) = 1/4$.
According to the finite-size scaling theory~\cite{challa_critical_1986, isakov_interplay_2003, zuo_scaling_2021},
for a finite-size system with linear size $L$, we have the scaling relation
\begin{equation} \label{eq:data-collapse}
      P_s L^{\eta(T_c)-2} = f\left(L^{-1}\mathrm{exp}\left(at_r^{-1/2}\right)\right),
      \quad t_r \to 0^+,
\end{equation}
where $f$ is a universal function. This relation gives us the freedom to appropriately choose $a$, $\eta$, and $T_c$
to make the plot of $P_s L^{\eta(T_c)-2}$ versus $L^{-1}\mathrm{exp}\left(at_r^{-1/2}\right)$ collapse into a universal curve.

\begin{figure}[htbp]
      \includegraphics[width=0.9\columnwidth]{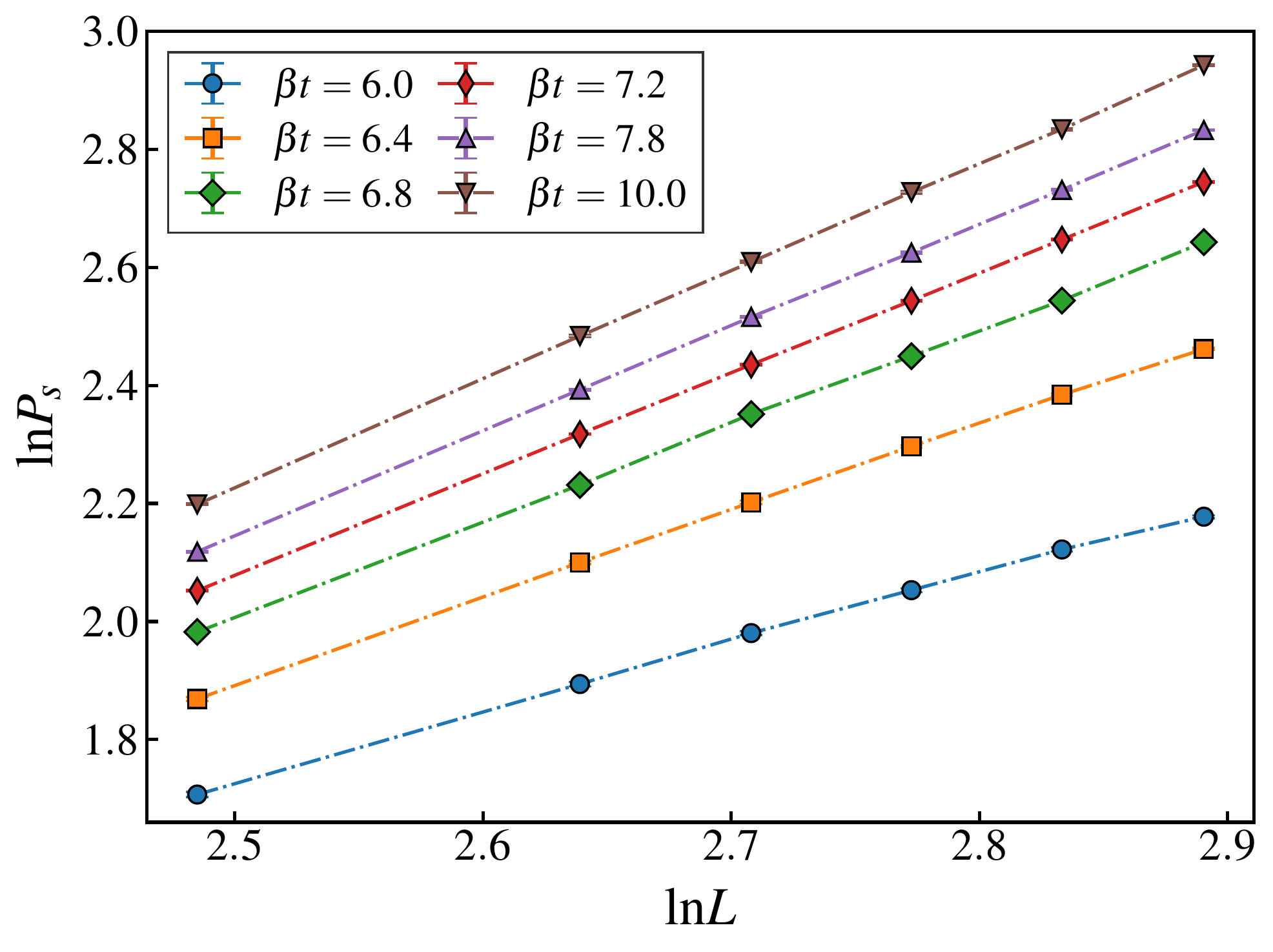}
      \caption{
            \label{fig:locate-Tc}
            Doubly logarithmic plot of the correlation function $P_s$ versus system size $L$ at various temperatures.
            Error bars are much smaller than symbol sizes.
      }
\end{figure}

Note that at the critical temperature $T_c$, we can conclude from Eq.~(\ref{eq:data-collapse}) that
\begin{equation} \label{eq:finite-size-scaling}
      P_s(L) \propto L^{2-\eta(T_c)}.
\end{equation}
Using Eq.~(\ref{eq:finite-size-scaling}), it is convenient to first locate $T_c$ by measuring $P_s$ for a range of temperature and lattice sizes.
Figure~\ref{fig:locate-Tc} shows the doubly logarithmic plot of the pairing $P_s$ as a function of system size $L$ at different temperatures.
Hereafter, we fix the value of interaction strength $|U|=4t$ and the filling number $\left\langle n\right\rangle=0.8$, 
which is around the optimal doping as indicated in Ref.~\cite{paiva_critical_2004}
(also note that we have set $t=1$ as the unit of energy).
The size of the lattice varies from $12\times12$ to $18\times18$.
As the inverse temperature $\beta$ increases from 6.0 to 10.0, the slope grows from $1.17\pm0.02$ to $1.83\pm0.01$.
According to Eq.~(\ref{eq:finite-size-scaling}), the slope should equal to 7/4 at the critical point.
This gives us the estimation of the transition temperature $\beta_c=7.75\pm0.10$.

\begin{figure}[htbp]
      \includegraphics[width=0.9\columnwidth]{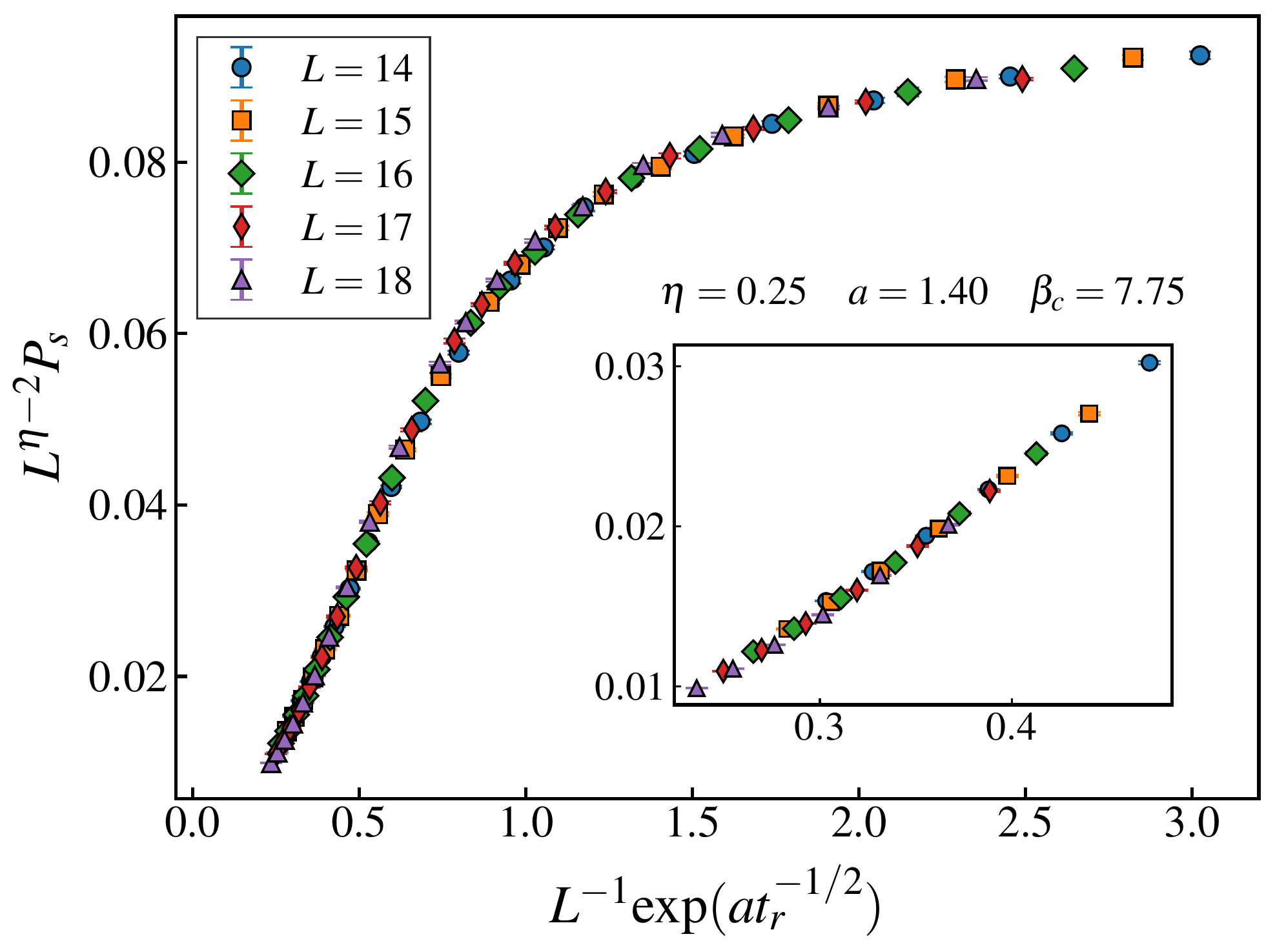}
      \caption{
            \label{fig:data-collapse}
            Data collapse of the correlation function $P_s(T,L)$.
            The inset is the zoom-in of the lower left corner of the main figure.
            The data points collapse quite well under the fitting parameters
            $a=1.40\pm0.10$ and $\beta_c=7.75\pm0.10$ (we have fixed $\eta=0.25$ for the BKT transition).
      }
\end{figure}

The $T$ dependence of the correlation length $\xi(T)$ can then be obtained using data collapse as introduced in Eq.~(\ref{eq:data-collapse}).
In Fig.~\ref{fig:data-collapse}, we show the data collapse of the pairing correlation function $P_s(T,L)$.
The data points fall into a universal function with the following set of fitting parameters:
$a=1.40\pm0.10$ and $\beta_c=7.75\pm0.10$ (we have fixed $\eta=0.25$ for the BKT transition).
Hence, the temperature dependence of the correlation length $\xi(T)$
is determined by Eq.~(\ref{eq:scaling}) with $a$ and $T_c$ given above, up to a $T$-independent prefactor.

\subsection{Pseudogap and the role of phase fluctuations}
The pseudogap phase is explicitly identified via solving the density of states $D(\omega)$
from the imaginary-time correlation function $D(\tau)=\frac{1}{L^2} \mathrm{Tr}\hat{G}(\tau)$ computed by DQMC.
The fermionic spectrum in real frequency is related to the correction function in imaginary time through an integral transform:
\begin{equation} \label{eq:integral-transform}
      D(\tau) = \int_{-\infty}^{\infty} \mathrm{d}\omega\ \frac{e^{-\tau\omega}}{1+e^{-\beta\omega}} D(\omega).
\end{equation}
Generally speaking, the extraction of $D(\omega)$ from $D(\tau)$ is numerically unstable
especially when the input correlation functions are noisy.
The SAC~\cite{sandvik_stochastic_1998,beach_identifying_2004,shao_nearly_2017,shao_progress_2023} method is thus proposed to hopefully find the most possible spectrum by carrying out a Monte Carlo simulation and simulated annealing procedure
(see detailed discussions in the Appendix).

\begin{figure}[htbp]
      \subfloat{
            \label{fig:dqmc-dos}
            \begin{overpic}[width=0.9\columnwidth]{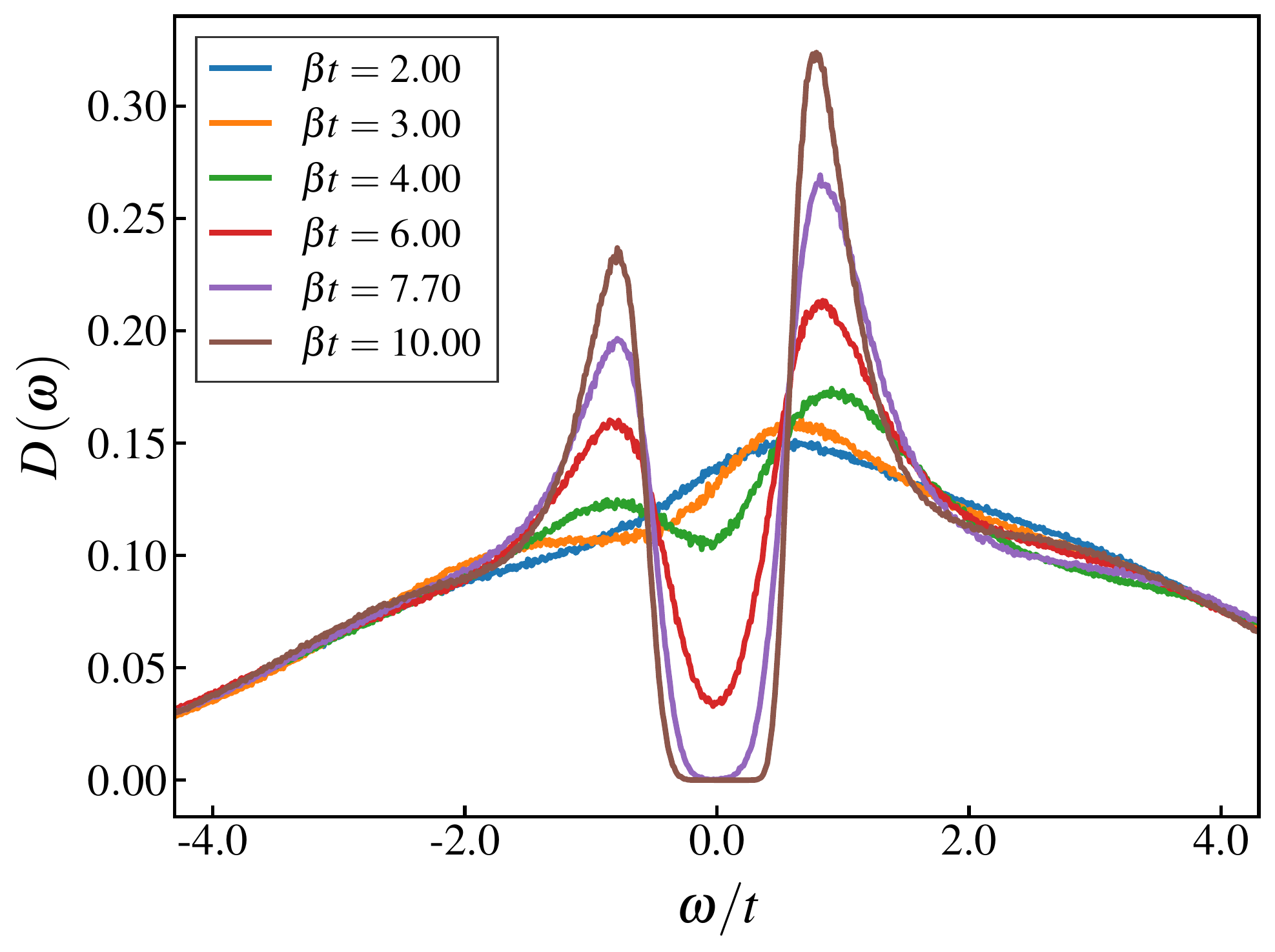}
                  \put(88,65){\large(a)}
            \end{overpic}
      }\\
      \subfloat{
            \label{fig:dqmc-spectral-functions}
            \begin{overpic}[width=0.9\columnwidth]{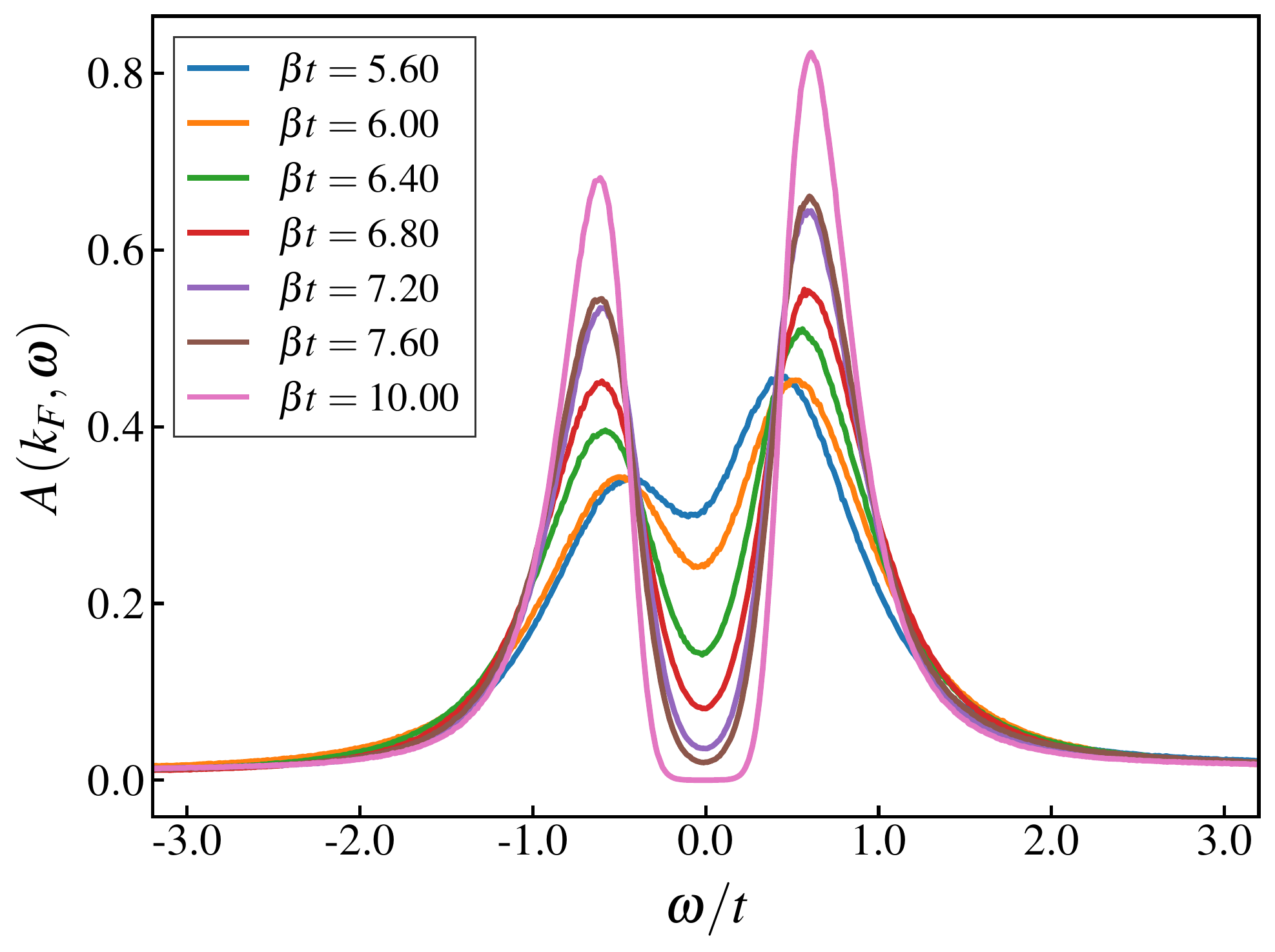}
                  \put(88,65){\large(b)}
            \end{overpic}
      }
      \caption{
            \label{fig:dqmc-results}
            (a) Density of state $D(\omega)$ at different temperatures
            for $|U|=4t$ and $\left\langle n\right\rangle=0.8$ on a $18\times18$ lattice.
            A smooth evolution of the superconducting gap is observed above the critical temperature.
            (b) Fermionic spectral function $A(k=k_F,\omega)$ at different temperatures.
            The momentum point is chosen as the intersection of the $\Delta$ line and the Fermi surface.
      }
\end{figure}

As a result, the recovered DOS is illustrated in Fig.~\ref{fig:dqmc-dos}.
Above the critical temperature $T_c$, the gap is gradually filled up
and finally closed at a temperature higher than 0.25.
This explicitly confirms the existence of the pseudogap phase in the intermediate-coupling regime of the attractive Hubbard model on a square lattice.

This pseudogap behavior can also be observed in the fermionic spectral functions,
which are obtained using a method similar to that used for getting the DOS.
We show in Fig.~\ref{fig:dqmc-spectral-functions} the spectral weights on the Fermi surface
at different temperatures.
Compared with our analysis in Eqs.~(\ref{eq:spectral-function}) and~(\ref{eq:half-width-relation}),
a qualitative difference is that even below the transition temperature $T_c$,
where the correlation length tends to infinity,
the two spectral peaks still possess a finite half-width.
This motivates us to introduce an intrinsic correlation length $\xi_0$,
which takes into account other effects beyond the phase fluctuations of the pairing order parameter, such as electron interactions, to the width of quasiparticle peaks:
\begin{equation}
      \Delta\omega = \frac{v_F}{2}\left(\xi^{-1}+{\xi_0^{-1}}\right).
\end{equation}

\begin{figure}[htbp]
      \includegraphics[width=0.9\columnwidth]{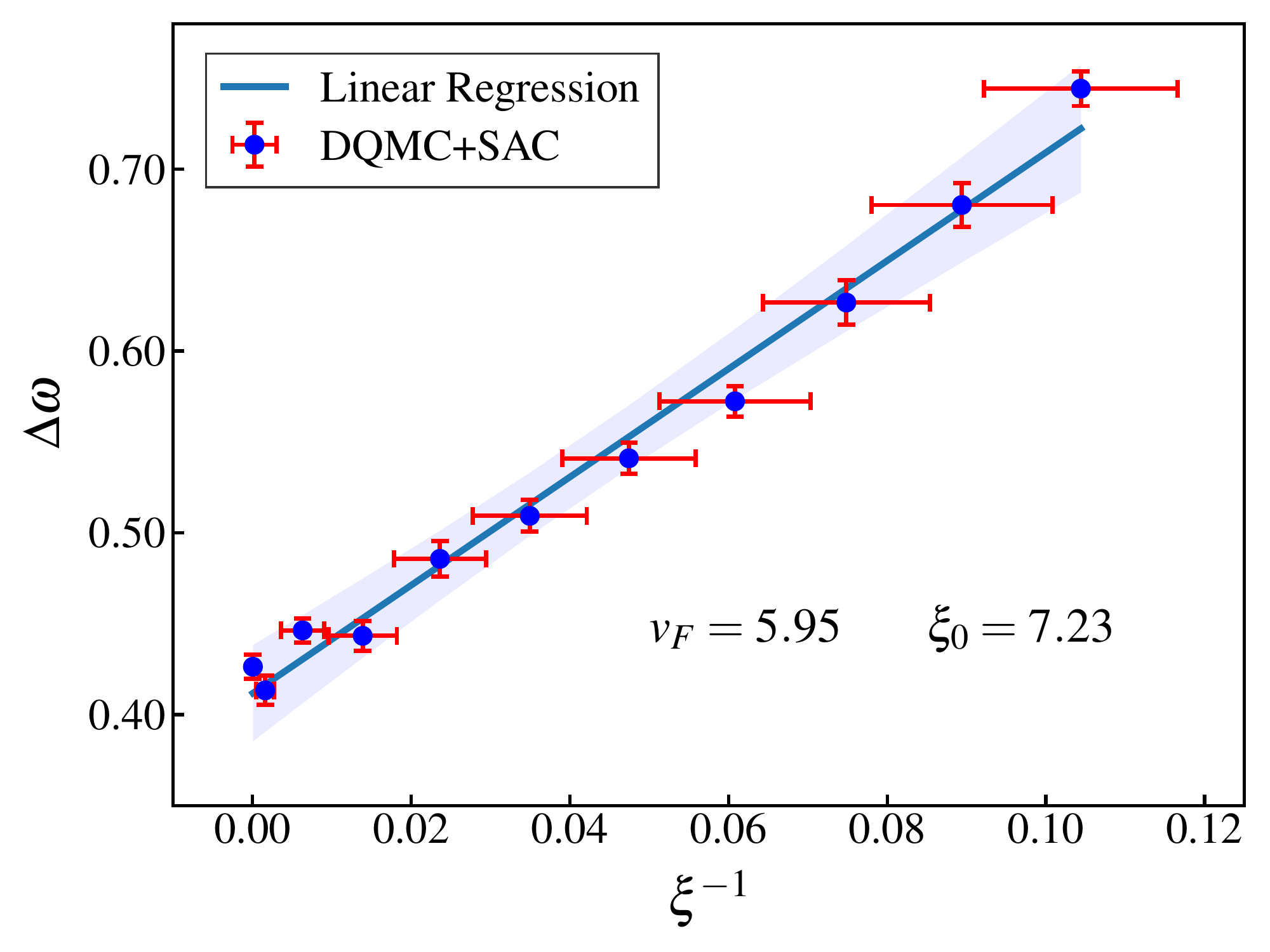}
      \caption{
            \label{fig:half-width-fit}
            Half-width $\Delta\omega$ of the spectral weight on the Fermi surface
            versus the inversion of the correlation length $\xi^{-1}$.
            The solid line is the result of linear fitting, and the shaded region represents the $3\sigma$ confidence interval of the regression.
      }
\end{figure}

The fitting of the half-width data and the correlation length is shown in Fig.~\ref{fig:half-width-fit}.
The half-width of spectral peaks is directly read from Fig.~\ref{fig:dqmc-spectral-functions},
and we estimate the error bars of the half-width by dividing the imaginary-time Green's functions measured by DQMC into 30 groups
and extracting the spectral functions using SAC independently.

Considering the numerical accuracy of both $\Delta\omega$ and $\xi$,
we conclude that the linear dependence of $\Delta\omega$ on $\xi$ is satisfied in the pseudogap regime.
With the inverse temperature ranging from 5.6 to 7.6, the magnitude of the correlation length rises from $10^{1}$ to $10^{4}$, which is compatible with our assumption of $k_F\xi\gg1$ in Eqs.~(\ref{eq:self-energy-approx}) and~(\ref{eq:half-width-relation}).
It can also be inferred from Fig.~\ref{fig:dqmc-spectral-functions} that
the superconducting gap $\Delta_0$ is around 0.60 for temperatures close to the critical point.
If we neglect the temperature dependence of $\Delta_0$ since we focus on the vicinity where the phase fluctuations count,
we can then acquire an estimation of the superconducting coherence length $\xi_\mathrm{BCS}\sim3.2$.
The above analysis justifies our perturbative expansion in Eq.~(\ref{eq:perturbative-expansion}) since the ratio $\Delta_0/8t$ is small.

In summary, we conclude that in the intermediate-coupling regime of the attractive Hubbard model,
the phase fluctuations above $T_c$ give birth to the pseudogap phenomenon,
where the smooth evolution of the superconducting gap occurs in the single-particle spectrum.
This is consistent with the theory developed in Sec.~\ref{sec:theory} that
in the presence of phase fluctuations the BCS quasiparticles acquire a finite lifetime and meanwhile the two spectral peaks get broadened.
At high temperatures, this further leads to the mergence of two BCS spectral peaks, and eventually the gap is closed.
Within the accuracy that we can achieve,
the half-width of spectral peaks on the Fermi surface is proportional to the ratio $\xi_\mathrm{BCS}/\xi$,
as predicted in Eq.~(\ref{eq:half-width-relation}).
\section{Conclusion}

In this paper, we consider the effect of
the phase fluctuations on the normal state of $s$-wave superconductors.
We start with a low-energy effective theory with electrons
coupled to a static superconducting pairing field,
which has a uniform magnitude and a randomly distributed phase.
We apply the technique of disorder averaging
to treat the contributions of the fluctuating phases,
and we obtain the self-energy corrections to the quasiparticles
in a perturbative manner.
We find that the phase fluctuations naturally provide
a well-defined imaginary part to the Green's function.
An intuitive relation then follows that
the broadening of the single-particle spectrum on the Fermi surface
is governed by the ratio of the superconducting coherence length $\xi_\mathrm{BCS}$
and the correlation length $\xi$.
This is supposed to be a universal relation regardless of the microscopic theory,
e.g., the mechanism of the superconducting pairing
or the line shape of the correlation functions.
The predictions given above are verified
by our DQMC simulations of the attractive-$U$ Hubbard model on the square lattice.

It will be interesting to see if the theoretical results hold in other models of 2D superconductors that can be simulated numerically.
Furthermore, the theoretical analysis in Sec.~\ref{sec:theory} only includes the leading term in the perturbation theory and, therefore, only works when the pairing gap is small.
It will be interesting to explore the effects of higher-order terms in the perturbative expansion.
We leave this to future works.

\begin{acknowledgments}
  The authors thank Chuang Chen, Yuan Da Liao, Yunchao Hao, and Kai Sun for invaluable discussions.
  Y.Q. is supported by the {\color{light-red} National Natural Science Foundation of China} (NSFC) through Grants No. {\color{dark-green} 11874115} and No. {\color{dark-green} 12174068}.
\end{acknowledgments}

\appendix
\section*{Appendix: Stochastic analytic continuation}

In this paper, we apply the stochastic analytic continuation (SAC)~\cite{sandvik_stochastic_1998,beach_identifying_2004,shao_nearly_2017,shao_progress_2023} method
to obtain the real-frequency spectral function from the imaginary-time correlation measured in QMC.
A brief introduction to the SAC method is summarized below.

In general, the numerical analytic continuation problem is formulated as solving the inverse of the following integral transform:
\begin{equation} \label{eq:general-integral-transform}
    G(\tau) = \int_{-\infty}^{\infty} \mathrm{d}\omega A(\omega) K(\tau,\omega),
\end{equation}
where $K(\tau,\omega)$ is the integral kernel, $A(\omega)$ is the spectral function to be solved, and $G(\tau)$ is the imaginary-time correlation sampled in QMC with certain statistical errors.
The kernel $K(\tau,\omega)$ depends on both temperature and the statistics of particles, and in our case of the fermionic system, we have $K(\tau,\omega)=e^{-\tau\omega}/\left(1+e^{-\beta\omega}\right)$ as implied in Eq.~(\ref{eq:integral-transform}).

Practically, the QMC simulation gives us a statistical estimate $\bar{G}_i\equiv\bar{G}(\tau_i)$ of the exact correlation $G(\tau)$ for a set of imaginary-time points $\{\tau_i\}$ for $i=1,...,N_\tau$.
The statistical errors of $\bar{G}(\tau_i)$ are described by the covariance matrix, which is given by
\begin{equation}
    C_{ij} = \frac{1}{N_B\left(N_B-1\right)} \sum_{b=1}^{N_B} \left(G^b_i-\bar{G}_i\right) \left(G^b_j-\bar{G}_j\right),
\end{equation}
where $G^b_i$ is the measured correlation at $\tau_i$ averaged from the QMC samples in a certain bin, and $N_B$ is the number of bins.
Note that the off-diagonal elements of the covariance matrix correspond to the correlation between different imaginary-time points, and the diagonal ones are exactly the square of the standard deviation of $\bar{G}_i$.

In order to numerically recover the spectral function, one should first parametrize $A(\omega)$ in an efficient and suitable way.
A widely used scheme is to represent $A(\omega)$ by a large number of $\delta$ functions:
\begin{equation}
    A(\omega) = \sum_{i=1}^{N_\omega} a_i \delta\left(\omega-\omega_i\right),
\end{equation}
with $\{a_i,\omega_i\}$ being the parameters to be sampled.
Since the fermionic spectral function is normalized, it is convenient to fix the amplitude $a_i \equiv a$ for all $i$ and sample over the locations of $\delta$ functions $\{\omega_i\}$.
Given certain configurations of $\delta$ functions, the fitted correlation $\tilde{G}_i$ can be computed according to Eq.~(\ref{eq:general-integral-transform})
and we further define the goodness of fit as
\begin{equation}
    \chi^2 = \sum_{i,j=1}^{N_\tau} \left(\tilde{G}_i-\bar{G}_i\right) C_{ij}^{-1} \left(\tilde{G}_j-\bar{G}_j\right),
\end{equation}
which quantifies the closeness of the fitted correlation to the QMC-measured one.

The key idea of SAC is to perform a Metropolis-Monte Carlo simulation in the configuration space $\{a_i,\omega_i\}$, and the weight of a certain spectrum obeys a Boltzmann distribution:
\begin{equation}
    W\left(\{a_i,\omega_i\}\right) \sim \exp\left(-\frac{\chi^2}{2\Theta}\right),
\end{equation}
where $\Theta$ is a fictitious temperature to balance the minimization of $\left\langle\chi^2\right\rangle$ and thermal fluctuations.
We also apply a simulated annealing process, which adiabatically tunes down $\Theta$ from a sufficiently high value, to avoid getting trapped into any configurations near local minima.
Once the optimal $\Theta$ is found, all statistically acceptable configurations are averaged to obtain the final spectrum.

\bibliographystyle{apsrev4-2}
\bibliography{reference}

\begin{thebibliography}{51}%
\makeatletter
\providecommand \@ifxundefined [1]{%
 \@ifx{#1\undefined}
}%
\providecommand \@ifnum [1]{%
 \ifnum #1\expandafter \@firstoftwo
 \else \expandafter \@secondoftwo
 \fi
}%
\providecommand \@ifx [1]{%
 \ifx #1\expandafter \@firstoftwo
 \else \expandafter \@secondoftwo
 \fi
}%
\providecommand \natexlab [1]{#1}%
\providecommand \enquote  [1]{``#1''}%
\providecommand \bibnamefont  [1]{#1}%
\providecommand \bibfnamefont [1]{#1}%
\providecommand \citenamefont [1]{#1}%
\providecommand \href@noop [0]{\@secondoftwo}%
\providecommand \href [0]{\begingroup \@sanitize@url \@href}%
\providecommand \@href[1]{\@@startlink{#1}\@@href}%
\providecommand \@@href[1]{\endgroup#1\@@endlink}%
\providecommand \@sanitize@url [0]{\catcode `\\12\catcode `\$12\catcode `\&12\catcode `\#12\catcode `\^12\catcode `\_12\catcode `\%12\relax}%
\providecommand \@@startlink[1]{}%
\providecommand \@@endlink[0]{}%
\providecommand \url  [0]{\begingroup\@sanitize@url \@url }%
\providecommand \@url [1]{\endgroup\@href {#1}{\urlprefix }}%
\providecommand \urlprefix  [0]{URL }%
\providecommand \Eprint [0]{\href }%
\providecommand \doibase [0]{https://doi.org/}%
\providecommand \selectlanguage [0]{\@gobble}%
\providecommand \bibinfo  [0]{\@secondoftwo}%
\providecommand \bibfield  [0]{\@secondoftwo}%
\providecommand \translation [1]{[#1]}%
\providecommand \BibitemOpen [0]{}%
\providecommand \bibitemStop [0]{}%
\providecommand \bibitemNoStop [0]{.\EOS\space}%
\providecommand \EOS [0]{\spacefactor3000\relax}%
\providecommand \BibitemShut  [1]{\csname bibitem#1\endcsname}%
\let\auto@bib@innerbib\@empty
\bibitem [{\citenamefont {Bednorz}\ and\ \citenamefont {M{\"u}ller}(1986)}]{Cuprate1986}%
  \BibitemOpen
  \bibfield  {author} {\bibinfo {author} {\bibfnamefont {J.~G.}\ \bibnamefont {Bednorz}}\ and\ \bibinfo {author} {\bibfnamefont {K.~A.}\ \bibnamefont {M{\"u}ller}},\ }\href {https://doi.org/10.1007/BF01303701} {\bibfield  {journal} {\bibinfo  {journal} {Z. Physik B - Condensed Matter}\ }\textbf {\bibinfo {volume} {64}},\ \bibinfo {pages} {189} (\bibinfo {year} {1986})}\BibitemShut {NoStop}%
\bibitem [{\citenamefont {Tsuei}\ and\ \citenamefont {Kirtley}(2000)}]{TsueiRMP2000}%
  \BibitemOpen
  \bibfield  {author} {\bibinfo {author} {\bibfnamefont {C.~C.}\ \bibnamefont {Tsuei}}\ and\ \bibinfo {author} {\bibfnamefont {J.~R.}\ \bibnamefont {Kirtley}},\ }\href {https://doi.org/10.1103/RevModPhys.72.969} {\bibfield  {journal} {\bibinfo  {journal} {Rev. Mod. Phys.}\ }\textbf {\bibinfo {volume} {72}},\ \bibinfo {pages} {969} (\bibinfo {year} {2000})}\BibitemShut {NoStop}%
\bibitem [{\citenamefont {Lee}\ \emph {et~al.}(2006)\citenamefont {Lee}, \citenamefont {Nagaosa},\ and\ \citenamefont {Wen}}]{LNW-HighTc-RMP2006}%
  \BibitemOpen
  \bibfield  {author} {\bibinfo {author} {\bibfnamefont {P.~A.}\ \bibnamefont {Lee}}, \bibinfo {author} {\bibfnamefont {N.}~\bibnamefont {Nagaosa}},\ and\ \bibinfo {author} {\bibfnamefont {X.-G.}\ \bibnamefont {Wen}},\ }\href {https://doi.org/10.1103/RevModPhys.78.17} {\bibfield  {journal} {\bibinfo  {journal} {Rev. Mod. Phys.}\ }\textbf {\bibinfo {volume} {78}},\ \bibinfo {pages} {17} (\bibinfo {year} {2006})}\BibitemShut {NoStop}%
\bibitem [{\citenamefont {Keimer}\ \emph {et~al.}(2015)\citenamefont {Keimer}, \citenamefont {Kivelson}, \citenamefont {Norman}, \citenamefont {Uchida},\ and\ \citenamefont {Zaanen}}]{Keimer2015Nature}%
  \BibitemOpen
  \bibfield  {author} {\bibinfo {author} {\bibfnamefont {B.}~\bibnamefont {Keimer}}, \bibinfo {author} {\bibfnamefont {S.~A.}\ \bibnamefont {Kivelson}}, \bibinfo {author} {\bibfnamefont {M.~R.}\ \bibnamefont {Norman}}, \bibinfo {author} {\bibfnamefont {S.}~\bibnamefont {Uchida}},\ and\ \bibinfo {author} {\bibfnamefont {J.}~\bibnamefont {Zaanen}},\ }\href {https://doi.org/10.1038/nature14165} {\bibfield  {journal} {\bibinfo  {journal} {Nature}\ }\textbf {\bibinfo {volume} {518}},\ \bibinfo {pages} {179} (\bibinfo {year} {2015})}\BibitemShut {NoStop}%
\bibitem [{\citenamefont {Proust}\ and\ \citenamefont {Taillefer}(2019)}]{Proust2018Review}%
  \BibitemOpen
  \bibfield  {author} {\bibinfo {author} {\bibfnamefont {C.}~\bibnamefont {Proust}}\ and\ \bibinfo {author} {\bibfnamefont {L.}~\bibnamefont {Taillefer}},\ }\href {https://doi.org/10.1146/annurev-conmatphys-031218-013210} {\bibfield  {journal} {\bibinfo  {journal} {Annu. Rev. Condens. Matter Phys.}\ }\textbf {\bibinfo {volume} {10}},\ \bibinfo {pages} {409} (\bibinfo {year} {2019})}\BibitemShut {NoStop}%
\bibitem [{\citenamefont {Arovas}\ \emph {et~al.}(2022)\citenamefont {Arovas}, \citenamefont {Berg}, \citenamefont {Kivelson},\ and\ \citenamefont {Raghu}}]{Arovas2022Review}%
  \BibitemOpen
  \bibfield  {author} {\bibinfo {author} {\bibfnamefont {D.~P.}\ \bibnamefont {Arovas}}, \bibinfo {author} {\bibfnamefont {E.}~\bibnamefont {Berg}}, \bibinfo {author} {\bibfnamefont {S.~A.}\ \bibnamefont {Kivelson}},\ and\ \bibinfo {author} {\bibfnamefont {S.}~\bibnamefont {Raghu}},\ }\href {https://doi.org/10.1146/annurev-conmatphys-031620-102024} {\bibfield  {journal} {\bibinfo  {journal} {Annu. Rev. Condens. Matter Phys.}\ }\textbf {\bibinfo {volume} {13}},\ \bibinfo {pages} {239} (\bibinfo {year} {2022})}\BibitemShut {NoStop}%
\bibitem [{\citenamefont {Timusk}\ and\ \citenamefont {Statt}(1999)}]{Timusk1999}%
  \BibitemOpen
  \bibfield  {author} {\bibinfo {author} {\bibfnamefont {T.}~\bibnamefont {Timusk}}\ and\ \bibinfo {author} {\bibfnamefont {B.}~\bibnamefont {Statt}},\ }\href {https://doi.org/10.1088/0034-4885/62/1/002} {\bibfield  {journal} {\bibinfo  {journal} {Rep. Prog. Phys.}\ }\textbf {\bibinfo {volume} {62}},\ \bibinfo {pages} {61} (\bibinfo {year} {1999})}\BibitemShut {NoStop}%
\bibitem [{\citenamefont {Vedeneev}(2021)}]{Vedeneev2021}%
  \BibitemOpen
  \bibfield  {author} {\bibinfo {author} {\bibfnamefont {S.~I.}\ \bibnamefont {Vedeneev}},\ }\href {https://doi.org/10.3367/UFNe.2020.12.038896} {\bibfield  {journal} {\bibinfo  {journal} {Phys.-Usp.}\ }\textbf {\bibinfo {volume} {64}},\ \bibinfo {pages} {890} (\bibinfo {year} {2021})}\BibitemShut {NoStop}%
\bibitem [{\citenamefont {Sobota}\ \emph {et~al.}(2021)\citenamefont {Sobota}, \citenamefont {He},\ and\ \citenamefont {Shen}}]{sobota_angle-resolved_2021}%
  \BibitemOpen
  \bibfield  {author} {\bibinfo {author} {\bibfnamefont {J.~A.}\ \bibnamefont {Sobota}}, \bibinfo {author} {\bibfnamefont {Y.}~\bibnamefont {He}},\ and\ \bibinfo {author} {\bibfnamefont {Z.-X.}\ \bibnamefont {Shen}},\ }\href {https://doi.org/10.1103/RevModPhys.93.025006} {\bibfield  {journal} {\bibinfo  {journal} {Rev. Mod. Phys.}\ }\textbf {\bibinfo {volume} {93}},\ \bibinfo {pages} {025006} (\bibinfo {year} {2021})}\BibitemShut {NoStop}%
\bibitem [{\citenamefont {Chakravarty}\ \emph {et~al.}(2001)\citenamefont {Chakravarty}, \citenamefont {Laughlin}, \citenamefont {Morr},\ and\ \citenamefont {Nayak}}]{chakravarty_hidden_2001}%
  \BibitemOpen
  \bibfield  {author} {\bibinfo {author} {\bibfnamefont {S.}~\bibnamefont {Chakravarty}}, \bibinfo {author} {\bibfnamefont {R.~B.}\ \bibnamefont {Laughlin}}, \bibinfo {author} {\bibfnamefont {D.~K.}\ \bibnamefont {Morr}},\ and\ \bibinfo {author} {\bibfnamefont {C.}~\bibnamefont {Nayak}},\ }\href {https://doi.org/10.1103/PhysRevB.63.094503} {\bibfield  {journal} {\bibinfo  {journal} {Phys. Rev. B}\ }\textbf {\bibinfo {volume} {63}},\ \bibinfo {pages} {094503} (\bibinfo {year} {2001})}\BibitemShut {NoStop}%
\bibitem [{\citenamefont {Gu}\ and\ \citenamefont {Weng}(2005)}]{GuUPG2005}%
  \BibitemOpen
  \bibfield  {author} {\bibinfo {author} {\bibfnamefont {Z.-C.}\ \bibnamefont {Gu}}\ and\ \bibinfo {author} {\bibfnamefont {Z.-Y.}\ \bibnamefont {Weng}},\ }\href {https://doi.org/10.1103/PhysRevB.72.104520} {\bibfield  {journal} {\bibinfo  {journal} {Phys. Rev. B}\ }\textbf {\bibinfo {volume} {72}},\ \bibinfo {pages} {104520} (\bibinfo {year} {2005})}\BibitemShut {NoStop}%
\bibitem [{\citenamefont {Varma}(2006)}]{varma_theory_2006}%
  \BibitemOpen
  \bibfield  {author} {\bibinfo {author} {\bibfnamefont {C.~M.}\ \bibnamefont {Varma}},\ }\href {https://doi.org/10.1103/PhysRevB.73.155113} {\bibfield  {journal} {\bibinfo  {journal} {Phys. Rev. B}\ }\textbf {\bibinfo {volume} {73}},\ \bibinfo {pages} {155113} (\bibinfo {year} {2006})}\BibitemShut {NoStop}%
\bibitem [{\citenamefont {Grilli}\ \emph {et~al.}(2009)\citenamefont {Grilli}, \citenamefont {Seibold}, \citenamefont {Di~Ciolo},\ and\ \citenamefont {Lorenzana}}]{grilli_fermi_2009}%
  \BibitemOpen
  \bibfield  {author} {\bibinfo {author} {\bibfnamefont {M.}~\bibnamefont {Grilli}}, \bibinfo {author} {\bibfnamefont {G.}~\bibnamefont {Seibold}}, \bibinfo {author} {\bibfnamefont {A.}~\bibnamefont {Di~Ciolo}},\ and\ \bibinfo {author} {\bibfnamefont {J.}~\bibnamefont {Lorenzana}},\ }\href {https://doi.org/10.1103/PhysRevB.79.125111} {\bibfield  {journal} {\bibinfo  {journal} {Phys. Rev. B}\ }\textbf {\bibinfo {volume} {79}},\ \bibinfo {pages} {125111} (\bibinfo {year} {2009})}\BibitemShut {NoStop}%
\bibitem [{\citenamefont {Li}\ \emph {et~al.}(2008)\citenamefont {Li}, \citenamefont {Balédent}, \citenamefont {Barišić}, \citenamefont {Cho}, \citenamefont {Fauqué}, \citenamefont {Sidis}, \citenamefont {Yu}, \citenamefont {Zhao}, \citenamefont {Bourges},\ and\ \citenamefont {Greven}}]{li_unusual_2008}%
  \BibitemOpen
  \bibfield  {author} {\bibinfo {author} {\bibfnamefont {Y.}~\bibnamefont {Li}}, \bibinfo {author} {\bibfnamefont {V.}~\bibnamefont {Balédent}}, \bibinfo {author} {\bibfnamefont {N.}~\bibnamefont {Barišić}}, \bibinfo {author} {\bibfnamefont {Y.}~\bibnamefont {Cho}}, \bibinfo {author} {\bibfnamefont {B.}~\bibnamefont {Fauqué}}, \bibinfo {author} {\bibfnamefont {Y.}~\bibnamefont {Sidis}}, \bibinfo {author} {\bibfnamefont {G.}~\bibnamefont {Yu}}, \bibinfo {author} {\bibfnamefont {X.}~\bibnamefont {Zhao}}, \bibinfo {author} {\bibfnamefont {P.}~\bibnamefont {Bourges}},\ and\ \bibinfo {author} {\bibfnamefont {M.}~\bibnamefont {Greven}},\ }\href {https://doi.org/10.1038/nature07251} {\bibfield  {journal} {\bibinfo  {journal} {Nature}\ }\textbf {\bibinfo {volume} {455}},\ \bibinfo {pages} {372} (\bibinfo {year} {2008})}\BibitemShut {NoStop}%
\bibitem [{\citenamefont {He}\ \emph {et~al.}(2011)\citenamefont {He}, \citenamefont {Hashimoto}, \citenamefont {Karapetyan}, \citenamefont {Koralek}, \citenamefont {Hinton}, \citenamefont {Testaud}, \citenamefont {Nathan}, \citenamefont {Yoshida}, \citenamefont {Yao}, \citenamefont {Tanaka}, \citenamefont {Meevasana}, \citenamefont {Moore}, \citenamefont {Lu}, \citenamefont {Mo}, \citenamefont {Ishikado}, \citenamefont {Eisaki}, \citenamefont {Hussain}, \citenamefont {Devereaux}, \citenamefont {Kivelson}, \citenamefont {Orenstein}, \citenamefont {Kapitulnik},\ and\ \citenamefont {Shen}}]{he_single-band_2011}%
  \BibitemOpen
  \bibfield  {author} {\bibinfo {author} {\bibfnamefont {R.-H.}\ \bibnamefont {He}}, \bibinfo {author} {\bibfnamefont {M.}~\bibnamefont {Hashimoto}}, \bibinfo {author} {\bibfnamefont {H.}~\bibnamefont {Karapetyan}}, \bibinfo {author} {\bibfnamefont {J.~D.}\ \bibnamefont {Koralek}}, \bibinfo {author} {\bibfnamefont {J.~P.}\ \bibnamefont {Hinton}}, \bibinfo {author} {\bibfnamefont {J.~P.}\ \bibnamefont {Testaud}}, \bibinfo {author} {\bibfnamefont {V.}~\bibnamefont {Nathan}}, \bibinfo {author} {\bibfnamefont {Y.}~\bibnamefont {Yoshida}}, \bibinfo {author} {\bibfnamefont {H.}~\bibnamefont {Yao}}, \bibinfo {author} {\bibfnamefont {K.}~\bibnamefont {Tanaka}}, \bibinfo {author} {\bibfnamefont {W.}~\bibnamefont {Meevasana}}, \bibinfo {author} {\bibfnamefont {R.~G.}\ \bibnamefont {Moore}}, \bibinfo {author} {\bibfnamefont {D.~H.}\ \bibnamefont {Lu}}, \bibinfo {author} {\bibfnamefont {S.-K.}\ \bibnamefont {Mo}}, \bibinfo {author} {\bibfnamefont {M.}~\bibnamefont {Ishikado}}, \bibinfo {author} {\bibfnamefont {H.}~\bibnamefont {Eisaki}}, \bibinfo {author} {\bibfnamefont {Z.}~\bibnamefont {Hussain}}, \bibinfo {author} {\bibfnamefont {T.~P.}\ \bibnamefont {Devereaux}}, \bibinfo {author} {\bibfnamefont {S.~A.}\ \bibnamefont {Kivelson}}, \bibinfo {author} {\bibfnamefont {J.}~\bibnamefont {Orenstein}}, \bibinfo {author} {\bibfnamefont {A.}~\bibnamefont {Kapitulnik}},\ and\ \bibinfo {author} {\bibfnamefont {Z.-X.}\ \bibnamefont {Shen}},\ }\href {https://doi.org/10.1126/science.1198415} {\bibfield  {journal} {\bibinfo  {journal} {Science}\ }\textbf {\bibinfo {volume} {331}},\ \bibinfo {pages} {1579} (\bibinfo {year} {2011})}\BibitemShut {NoStop}%
\bibitem [{\citenamefont {Chang}\ \emph {et~al.}(2012)\citenamefont {Chang}, \citenamefont {Blackburn}, \citenamefont {Holmes}, \citenamefont {Christensen}, \citenamefont {Larsen}, \citenamefont {Mesot}, \citenamefont {Liang}, \citenamefont {Bonn}, \citenamefont {Hardy}, \citenamefont {Watenphul}, \citenamefont {Zimmermann}, \citenamefont {Forgan},\ and\ \citenamefont {Hayden}}]{chang_direct_2012}%
  \BibitemOpen
  \bibfield  {author} {\bibinfo {author} {\bibfnamefont {J.}~\bibnamefont {Chang}}, \bibinfo {author} {\bibfnamefont {E.}~\bibnamefont {Blackburn}}, \bibinfo {author} {\bibfnamefont {A.~T.}\ \bibnamefont {Holmes}}, \bibinfo {author} {\bibfnamefont {N.~B.}\ \bibnamefont {Christensen}}, \bibinfo {author} {\bibfnamefont {J.}~\bibnamefont {Larsen}}, \bibinfo {author} {\bibfnamefont {J.}~\bibnamefont {Mesot}}, \bibinfo {author} {\bibfnamefont {R.}~\bibnamefont {Liang}}, \bibinfo {author} {\bibfnamefont {D.~A.}\ \bibnamefont {Bonn}}, \bibinfo {author} {\bibfnamefont {W.~N.}\ \bibnamefont {Hardy}}, \bibinfo {author} {\bibfnamefont {A.}~\bibnamefont {Watenphul}}, \bibinfo {author} {\bibfnamefont {M.~v.}\ \bibnamefont {Zimmermann}}, \bibinfo {author} {\bibfnamefont {E.~M.}\ \bibnamefont {Forgan}},\ and\ \bibinfo {author} {\bibfnamefont {S.~M.}\ \bibnamefont {Hayden}},\ }\href {https://doi.org/10.1038/nphys2456} {\bibfield  {journal} {\bibinfo  {journal} {Nat. Phys.}\ }\textbf {\bibinfo {volume} {8}},\ \bibinfo {pages} {871} (\bibinfo {year} {2012})}\BibitemShut {NoStop}%
\bibitem [{\citenamefont {Franz}(2007)}]{franz_importance_2007}%
  \BibitemOpen
  \bibfield  {author} {\bibinfo {author} {\bibfnamefont {M.}~\bibnamefont {Franz}},\ }\href {https://doi.org/10.1038/nphys739} {\bibfield  {journal} {\bibinfo  {journal} {Nat. Phys.}\ }\textbf {\bibinfo {volume} {3}},\ \bibinfo {pages} {686} (\bibinfo {year} {2007})}\BibitemShut {NoStop}%
\bibitem [{\citenamefont {Hashimoto}\ \emph {et~al.}(2014)\citenamefont {Hashimoto}, \citenamefont {Vishik}, \citenamefont {He}, \citenamefont {Devereaux},\ and\ \citenamefont {Shen}}]{hashimoto_energy_2014}%
  \BibitemOpen
  \bibfield  {author} {\bibinfo {author} {\bibfnamefont {M.}~\bibnamefont {Hashimoto}}, \bibinfo {author} {\bibfnamefont {I.~M.}\ \bibnamefont {Vishik}}, \bibinfo {author} {\bibfnamefont {R.-H.}\ \bibnamefont {He}}, \bibinfo {author} {\bibfnamefont {T.~P.}\ \bibnamefont {Devereaux}},\ and\ \bibinfo {author} {\bibfnamefont {Z.-X.}\ \bibnamefont {Shen}},\ }\href {https://doi.org/10.1038/nphys3009} {\bibfield  {journal} {\bibinfo  {journal} {Nat. Phys.}\ }\textbf {\bibinfo {volume} {10}},\ \bibinfo {pages} {483} (\bibinfo {year} {2014})}\BibitemShut {NoStop}%
\bibitem [{\citenamefont {Hashimoto}\ \emph {et~al.}(2015)\citenamefont {Hashimoto}, \citenamefont {Nowadnick}, \citenamefont {He}, \citenamefont {Vishik}, \citenamefont {Moritz}, \citenamefont {He}, \citenamefont {Tanaka}, \citenamefont {Moore}, \citenamefont {Lu}, \citenamefont {Yoshida}, \citenamefont {Ishikado}, \citenamefont {Sasagawa}, \citenamefont {Fujita}, \citenamefont {Ishida}, \citenamefont {Uchida}, \citenamefont {Eisaki}, \citenamefont {Hussain}, \citenamefont {Devereaux},\ and\ \citenamefont {Shen}}]{hashimoto_direct_2015}%
  \BibitemOpen
  \bibfield  {author} {\bibinfo {author} {\bibfnamefont {M.}~\bibnamefont {Hashimoto}}, \bibinfo {author} {\bibfnamefont {E.~A.}\ \bibnamefont {Nowadnick}}, \bibinfo {author} {\bibfnamefont {R.-H.}\ \bibnamefont {He}}, \bibinfo {author} {\bibfnamefont {I.~M.}\ \bibnamefont {Vishik}}, \bibinfo {author} {\bibfnamefont {B.}~\bibnamefont {Moritz}}, \bibinfo {author} {\bibfnamefont {Y.}~\bibnamefont {He}}, \bibinfo {author} {\bibfnamefont {K.}~\bibnamefont {Tanaka}}, \bibinfo {author} {\bibfnamefont {R.~G.}\ \bibnamefont {Moore}}, \bibinfo {author} {\bibfnamefont {D.}~\bibnamefont {Lu}}, \bibinfo {author} {\bibfnamefont {Y.}~\bibnamefont {Yoshida}}, \bibinfo {author} {\bibfnamefont {M.}~\bibnamefont {Ishikado}}, \bibinfo {author} {\bibfnamefont {T.}~\bibnamefont {Sasagawa}}, \bibinfo {author} {\bibfnamefont {K.}~\bibnamefont {Fujita}}, \bibinfo {author} {\bibfnamefont {S.}~\bibnamefont {Ishida}}, \bibinfo {author} {\bibfnamefont {S.}~\bibnamefont {Uchida}}, \bibinfo {author} {\bibfnamefont {H.}~\bibnamefont {Eisaki}}, \bibinfo {author} {\bibfnamefont {Z.}~\bibnamefont {Hussain}}, \bibinfo {author} {\bibfnamefont {T.~P.}\ \bibnamefont {Devereaux}},\ and\ \bibinfo {author} {\bibfnamefont {Z.-X.}\ \bibnamefont {Shen}},\ }\href {https://doi.org/10.1038/nmat4116} {\bibfield  {journal} {\bibinfo  {journal} {Nat. Mater.}\ }\textbf {\bibinfo {volume} {14}},\ \bibinfo {pages} {37} (\bibinfo {year} {2015})}\BibitemShut {NoStop}%
\bibitem [{\citenamefont {Kondo}\ \emph {et~al.}(2015)\citenamefont {Kondo}, \citenamefont {Malaeb}, \citenamefont {Ishida}, \citenamefont {Sasagawa}, \citenamefont {Sakamoto}, \citenamefont {Takeuchi}, \citenamefont {Tohyama},\ and\ \citenamefont {Shin}}]{kondo_point_2015}%
  \BibitemOpen
  \bibfield  {author} {\bibinfo {author} {\bibfnamefont {T.}~\bibnamefont {Kondo}}, \bibinfo {author} {\bibfnamefont {W.}~\bibnamefont {Malaeb}}, \bibinfo {author} {\bibfnamefont {Y.}~\bibnamefont {Ishida}}, \bibinfo {author} {\bibfnamefont {T.}~\bibnamefont {Sasagawa}}, \bibinfo {author} {\bibfnamefont {H.}~\bibnamefont {Sakamoto}}, \bibinfo {author} {\bibfnamefont {T.}~\bibnamefont {Takeuchi}}, \bibinfo {author} {\bibfnamefont {T.}~\bibnamefont {Tohyama}},\ and\ \bibinfo {author} {\bibfnamefont {S.}~\bibnamefont {Shin}},\ }\href {https://doi.org/10.1038/ncomms8699} {\bibfield  {journal} {\bibinfo  {journal} {Nat. Commun.}\ }\textbf {\bibinfo {volume} {6}},\ \bibinfo {pages} {7699} (\bibinfo {year} {2015})}\BibitemShut {NoStop}%
\bibitem [{\citenamefont {Chen}\ \emph {et~al.}(2019)\citenamefont {Chen}, \citenamefont {Hashimoto}, \citenamefont {He}, \citenamefont {Song}, \citenamefont {Xu}, \citenamefont {He}, \citenamefont {Devereaux}, \citenamefont {Eisaki}, \citenamefont {Lu}, \citenamefont {Zaanen},\ and\ \citenamefont {Shen}}]{chen_incoherent_2019}%
  \BibitemOpen
  \bibfield  {author} {\bibinfo {author} {\bibfnamefont {S.-D.}\ \bibnamefont {Chen}}, \bibinfo {author} {\bibfnamefont {M.}~\bibnamefont {Hashimoto}}, \bibinfo {author} {\bibfnamefont {Y.}~\bibnamefont {He}}, \bibinfo {author} {\bibfnamefont {D.}~\bibnamefont {Song}}, \bibinfo {author} {\bibfnamefont {K.-J.}\ \bibnamefont {Xu}}, \bibinfo {author} {\bibfnamefont {J.-F.}\ \bibnamefont {He}}, \bibinfo {author} {\bibfnamefont {T.~P.}\ \bibnamefont {Devereaux}}, \bibinfo {author} {\bibfnamefont {H.}~\bibnamefont {Eisaki}}, \bibinfo {author} {\bibfnamefont {D.-H.}\ \bibnamefont {Lu}}, \bibinfo {author} {\bibfnamefont {J.}~\bibnamefont {Zaanen}},\ and\ \bibinfo {author} {\bibfnamefont {Z.-X.}\ \bibnamefont {Shen}},\ }\href {https://doi.org/10.1126/science.aaw8850} {\bibfield  {journal} {\bibinfo  {journal} {Science}\ }\textbf {\bibinfo {volume} {366}},\ \bibinfo {pages} {1099} (\bibinfo {year} {2019})}\BibitemShut {NoStop}%
\bibitem [{\citenamefont {Kosterlitz}\ and\ \citenamefont {Thouless}(1973)}]{kosterlitz_ordering_1973}%
  \BibitemOpen
  \bibfield  {author} {\bibinfo {author} {\bibfnamefont {J.~M.}\ \bibnamefont {Kosterlitz}}\ and\ \bibinfo {author} {\bibfnamefont {D.~J.}\ \bibnamefont {Thouless}},\ }\href {https://doi.org/10.1088/0022-3719/6/7/010} {\bibfield  {journal} {\bibinfo  {journal} {Journal of Physics C: Solid State Physics}\ }\textbf {\bibinfo {volume} {6}},\ \bibinfo {pages} {1181} (\bibinfo {year} {1973})}\BibitemShut {NoStop}%
\bibitem [{\citenamefont {Kosterlitz}(1974)}]{kosterlitz_critical_1974}%
  \BibitemOpen
  \bibfield  {author} {\bibinfo {author} {\bibfnamefont {J.~M.}\ \bibnamefont {Kosterlitz}},\ }\href {https://doi.org/10.1088/0022-3719/7/6/005} {\bibfield  {journal} {\bibinfo  {journal} {Journal of Physics C: Solid State Physics}\ }\textbf {\bibinfo {volume} {7}},\ \bibinfo {pages} {1046} (\bibinfo {year} {1974})}\BibitemShut {NoStop}%
\bibitem [{\citenamefont {Xu}\ \emph {et~al.}(2000)\citenamefont {Xu}, \citenamefont {Ong}, \citenamefont {Wang}, \citenamefont {Kakeshita},\ and\ \citenamefont {Uchida}}]{xu_vortex-like_2000}%
  \BibitemOpen
  \bibfield  {author} {\bibinfo {author} {\bibfnamefont {Z.~A.}\ \bibnamefont {Xu}}, \bibinfo {author} {\bibfnamefont {N.~P.}\ \bibnamefont {Ong}}, \bibinfo {author} {\bibfnamefont {Y.}~\bibnamefont {Wang}}, \bibinfo {author} {\bibfnamefont {T.}~\bibnamefont {Kakeshita}},\ and\ \bibinfo {author} {\bibfnamefont {S.}~\bibnamefont {Uchida}},\ }\href {https://doi.org/10.1038/35020016} {\bibfield  {journal} {\bibinfo  {journal} {Nature}\ }\textbf {\bibinfo {volume} {406}},\ \bibinfo {pages} {486} (\bibinfo {year} {2000})}\BibitemShut {NoStop}%
\bibitem [{\citenamefont {Balci}\ \emph {et~al.}(2003)\citenamefont {Balci}, \citenamefont {Hill}, \citenamefont {Qazilbash},\ and\ \citenamefont {Greene}}]{balci_nernst_2003}%
  \BibitemOpen
  \bibfield  {author} {\bibinfo {author} {\bibfnamefont {H.}~\bibnamefont {Balci}}, \bibinfo {author} {\bibfnamefont {C.~P.}\ \bibnamefont {Hill}}, \bibinfo {author} {\bibfnamefont {M.~M.}\ \bibnamefont {Qazilbash}},\ and\ \bibinfo {author} {\bibfnamefont {R.~L.}\ \bibnamefont {Greene}},\ }\href {https://doi.org/10.1103/PhysRevB.68.054520} {\bibfield  {journal} {\bibinfo  {journal} {Phys. Rev. B}\ }\textbf {\bibinfo {volume} {68}},\ \bibinfo {pages} {054520} (\bibinfo {year} {2003})}\BibitemShut {NoStop}%
\bibitem [{\citenamefont {Wang}\ \emph {et~al.}(2006)\citenamefont {Wang}, \citenamefont {Li},\ and\ \citenamefont {Ong}}]{wang_nernst_2006}%
  \BibitemOpen
  \bibfield  {author} {\bibinfo {author} {\bibfnamefont {Y.}~\bibnamefont {Wang}}, \bibinfo {author} {\bibfnamefont {L.}~\bibnamefont {Li}},\ and\ \bibinfo {author} {\bibfnamefont {N.~P.}\ \bibnamefont {Ong}},\ }\href {https://doi.org/10.1103/PhysRevB.73.024510} {\bibfield  {journal} {\bibinfo  {journal} {Phys. Rev. B}\ }\textbf {\bibinfo {volume} {73}},\ \bibinfo {pages} {024510} (\bibinfo {year} {2006})}\BibitemShut {NoStop}%
\bibitem [{\citenamefont {Bergeal}\ \emph {et~al.}(2008)\citenamefont {Bergeal}, \citenamefont {Lesueur}, \citenamefont {Aprili}, \citenamefont {Faini}, \citenamefont {Contour},\ and\ \citenamefont {Leridon}}]{bergeal_pairing_2008}%
  \BibitemOpen
  \bibfield  {author} {\bibinfo {author} {\bibfnamefont {N.}~\bibnamefont {Bergeal}}, \bibinfo {author} {\bibfnamefont {J.}~\bibnamefont {Lesueur}}, \bibinfo {author} {\bibfnamefont {M.}~\bibnamefont {Aprili}}, \bibinfo {author} {\bibfnamefont {G.}~\bibnamefont {Faini}}, \bibinfo {author} {\bibfnamefont {J.~P.}\ \bibnamefont {Contour}},\ and\ \bibinfo {author} {\bibfnamefont {B.}~\bibnamefont {Leridon}},\ }\href {https://doi.org/10.1038/nphys1017} {\bibfield  {journal} {\bibinfo  {journal} {Nat. Phys.}\ }\textbf {\bibinfo {volume} {4}},\ \bibinfo {pages} {608} (\bibinfo {year} {2008})}\BibitemShut {NoStop}%
\bibitem [{\citenamefont {Kwon}\ and\ \citenamefont {Dorsey}(1999)}]{kwon_effect_1999}%
  \BibitemOpen
  \bibfield  {author} {\bibinfo {author} {\bibfnamefont {H.-J.}\ \bibnamefont {Kwon}}\ and\ \bibinfo {author} {\bibfnamefont {A.~T.}\ \bibnamefont {Dorsey}},\ }\href {https://doi.org/10.1103/PhysRevB.59.6438} {\bibfield  {journal} {\bibinfo  {journal} {Phys. Rev. B}\ }\textbf {\bibinfo {volume} {59}},\ \bibinfo {pages} {6438} (\bibinfo {year} {1999})}\BibitemShut {NoStop}%
\bibitem [{\citenamefont {Franz}\ and\ \citenamefont {Millis}(1998)}]{franz_phase_1998}%
  \BibitemOpen
  \bibfield  {author} {\bibinfo {author} {\bibfnamefont {M.}~\bibnamefont {Franz}}\ and\ \bibinfo {author} {\bibfnamefont {A.~J.}\ \bibnamefont {Millis}},\ }\href {https://doi.org/10.1103/PhysRevB.58.14572} {\bibfield  {journal} {\bibinfo  {journal} {Phys. Rev. B}\ }\textbf {\bibinfo {volume} {58}},\ \bibinfo {pages} {14572} (\bibinfo {year} {1998})}\BibitemShut {NoStop}%
\bibitem [{\citenamefont {Kwon}\ \emph {et~al.}(2001)\citenamefont {Kwon}, \citenamefont {Dorsey},\ and\ \citenamefont {Hirschfeld}}]{kwon_observability_2001}%
  \BibitemOpen
  \bibfield  {author} {\bibinfo {author} {\bibfnamefont {H.-J.}\ \bibnamefont {Kwon}}, \bibinfo {author} {\bibfnamefont {A.~T.}\ \bibnamefont {Dorsey}},\ and\ \bibinfo {author} {\bibfnamefont {P.~J.}\ \bibnamefont {Hirschfeld}},\ }\href {https://doi.org/10.1103/PhysRevLett.86.3875} {\bibfield  {journal} {\bibinfo  {journal} {Phys. Rev. Lett.}\ }\textbf {\bibinfo {volume} {86}},\ \bibinfo {pages} {3875} (\bibinfo {year} {2001})}\BibitemShut {NoStop}%
\bibitem [{\citenamefont {Franz}\ and\ \citenamefont {Tešanović}(2001)}]{franz_algebraic_2001}%
  \BibitemOpen
  \bibfield  {author} {\bibinfo {author} {\bibfnamefont {M.}~\bibnamefont {Franz}}\ and\ \bibinfo {author} {\bibfnamefont {Z.}~\bibnamefont {Tešanović}},\ }\href {https://doi.org/10.1103/PhysRevLett.87.257003} {\bibfield  {journal} {\bibinfo  {journal} {Phys. Rev. Lett.}\ }\textbf {\bibinfo {volume} {87}},\ \bibinfo {pages} {257003} (\bibinfo {year} {2001})}\BibitemShut {NoStop}%
\bibitem [{\citenamefont {Curty}\ and\ \citenamefont {Beck}(2003)}]{curty_thermodynamics_2003}%
  \BibitemOpen
  \bibfield  {author} {\bibinfo {author} {\bibfnamefont {P.}~\bibnamefont {Curty}}\ and\ \bibinfo {author} {\bibfnamefont {H.}~\bibnamefont {Beck}},\ }\href {https://doi.org/10.1103/PhysRevLett.91.257002} {\bibfield  {journal} {\bibinfo  {journal} {Phys. Rev. Lett.}\ }\textbf {\bibinfo {volume} {91}},\ \bibinfo {pages} {257002} (\bibinfo {year} {2003})}\BibitemShut {NoStop}%
\bibitem [{\citenamefont {Mahan}(2000)}]{mahan_many_2000}%
  \BibitemOpen
  \bibfield  {author} {\bibinfo {author} {\bibfnamefont {G.~D.}\ \bibnamefont {Mahan}},\ }\href {https://doi.org/10.1007/978-1-4757-5714-9} {\emph {\bibinfo {title} {Many-{Particle} {Physics}}}},\ \bibinfo {edition} {3rd}\ ed.\ (\bibinfo  {publisher} {Springer, New York},\ \bibinfo {year} {2000})\BibitemShut {NoStop}%
\bibitem [{\citenamefont {Nagai}\ \emph {et~al.}(2020)\citenamefont {Nagai}, \citenamefont {Qi}, \citenamefont {Isobe}, \citenamefont {Kozii},\ and\ \citenamefont {Fu}}]{nagai_dmft_2020}%
  \BibitemOpen
  \bibfield  {author} {\bibinfo {author} {\bibfnamefont {Y.}~\bibnamefont {Nagai}}, \bibinfo {author} {\bibfnamefont {Y.}~\bibnamefont {Qi}}, \bibinfo {author} {\bibfnamefont {H.}~\bibnamefont {Isobe}}, \bibinfo {author} {\bibfnamefont {V.}~\bibnamefont {Kozii}},\ and\ \bibinfo {author} {\bibfnamefont {L.}~\bibnamefont {Fu}},\ }\href {https://doi.org/10.1103/PhysRevLett.125.227204} {\bibfield  {journal} {\bibinfo  {journal} {Phys. Rev. Lett.}\ }\textbf {\bibinfo {volume} {125}},\ \bibinfo {pages} {227204} (\bibinfo {year} {2020})}\BibitemShut {NoStop}%
\bibitem [{\citenamefont {Annett}(2003)}]{annett_superconductivity_2003}%
  \BibitemOpen
  \bibfield  {author} {\bibinfo {author} {\bibfnamefont {J.~F.}\ \bibnamefont {Annett}},\ }\href {https://global.oup.com/academic/product/superconductivity-superfluids-and-condensates-9780198507567?q=9780198507567&cc=ca&lang=en#} {\emph {\bibinfo {title} {Superconductivity, {Superﬂuids} and {Condensates}}}},\ \bibinfo {edition} {1st}\ ed.\ (\bibinfo  {publisher} {Oxford University, Oxford},\ \bibinfo {year} {2003})\BibitemShut {NoStop}%
\bibitem [{\citenamefont {Singh}\ \emph {et~al.}(2022)\citenamefont {Singh}, \citenamefont {Kadge}, \citenamefont {Bang},\ and\ \citenamefont {Majumdar}}]{singh_fermi_2022}%
  \BibitemOpen
  \bibfield  {author} {\bibinfo {author} {\bibfnamefont {D.~K.}\ \bibnamefont {Singh}}, \bibinfo {author} {\bibfnamefont {S.}~\bibnamefont {Kadge}}, \bibinfo {author} {\bibfnamefont {Y.}~\bibnamefont {Bang}},\ and\ \bibinfo {author} {\bibfnamefont {P.}~\bibnamefont {Majumdar}},\ }\href {https://doi.org/10.1103/PhysRevB.105.054501} {\bibfield  {journal} {\bibinfo  {journal} {Phys. Rev. B}\ }\textbf {\bibinfo {volume} {105}},\ \bibinfo {pages} {054501} (\bibinfo {year} {2022})}\BibitemShut {NoStop}%
\bibitem [{\citenamefont {Wu}\ and\ \citenamefont {Zhang}(2005)}]{wu_sufficient_2005}%
  \BibitemOpen
  \bibfield  {author} {\bibinfo {author} {\bibfnamefont {C.}~\bibnamefont {Wu}}\ and\ \bibinfo {author} {\bibfnamefont {S.-C.}\ \bibnamefont {Zhang}},\ }\href {https://doi.org/10.1103/PhysRevB.71.155115} {\bibfield  {journal} {\bibinfo  {journal} {Phys. Rev. B}\ }\textbf {\bibinfo {volume} {71}},\ \bibinfo {pages} {155115} (\bibinfo {year} {2005})}\BibitemShut {NoStop}%
\bibitem [{\citenamefont {Moreo}\ and\ \citenamefont {Scalapino}(1991)}]{moreo_two-dimensional_1991}%
  \BibitemOpen
  \bibfield  {author} {\bibinfo {author} {\bibfnamefont {A.}~\bibnamefont {Moreo}}\ and\ \bibinfo {author} {\bibfnamefont {D.~J.}\ \bibnamefont {Scalapino}},\ }\href {https://doi.org/10.1103/PhysRevLett.66.946} {\bibfield  {journal} {\bibinfo  {journal} {Phys. Rev. Lett.}\ }\textbf {\bibinfo {volume} {66}},\ \bibinfo {pages} {946} (\bibinfo {year} {1991})}\BibitemShut {NoStop}%
\bibitem [{\citenamefont {Paiva}\ \emph {et~al.}(2004)\citenamefont {Paiva}, \citenamefont {dos Santos}, \citenamefont {Scalettar},\ and\ \citenamefont {Denteneer}}]{paiva_critical_2004}%
  \BibitemOpen
  \bibfield  {author} {\bibinfo {author} {\bibfnamefont {T.}~\bibnamefont {Paiva}}, \bibinfo {author} {\bibfnamefont {R.~R.}\ \bibnamefont {dos Santos}}, \bibinfo {author} {\bibfnamefont {R.~T.}\ \bibnamefont {Scalettar}},\ and\ \bibinfo {author} {\bibfnamefont {P.~J.~H.}\ \bibnamefont {Denteneer}},\ }\href {https://doi.org/10.1103/PhysRevB.69.184501} {\bibfield  {journal} {\bibinfo  {journal} {Phys. Rev. B}\ }\textbf {\bibinfo {volume} {69}},\ \bibinfo {pages} {184501} (\bibinfo {year} {2004})}\BibitemShut {NoStop}%
\bibitem [{\citenamefont {Vilk}\ \emph {et~al.}(1998)\citenamefont {Vilk}, \citenamefont {Allen}, \citenamefont {Touchette}, \citenamefont {Moukouri}, \citenamefont {Chen},\ and\ \citenamefont {Tremblay}}]{vilk_attractive_1998}%
  \BibitemOpen
  \bibfield  {author} {\bibinfo {author} {\bibfnamefont {Y.~M.}\ \bibnamefont {Vilk}}, \bibinfo {author} {\bibfnamefont {S.}~\bibnamefont {Allen}}, \bibinfo {author} {\bibfnamefont {H.}~\bibnamefont {Touchette}}, \bibinfo {author} {\bibfnamefont {S.}~\bibnamefont {Moukouri}}, \bibinfo {author} {\bibfnamefont {L.}~\bibnamefont {Chen}},\ and\ \bibinfo {author} {\bibfnamefont {A.~M.~S.}\ \bibnamefont {Tremblay}},\ }\href {https://doi.org/10.1016/S0022-3697(98)00129-2} {\bibfield  {journal} {\bibinfo  {journal} {Journal of Physics and Chemistry of Solids}\ }\textbf {\bibinfo {volume} {59}},\ \bibinfo {pages} {1873} (\bibinfo {year} {1998})}\BibitemShut {NoStop}%
\bibitem [{\citenamefont {Randeria}\ \emph {et~al.}(1992)\citenamefont {Randeria}, \citenamefont {Trivedi}, \citenamefont {Moreo},\ and\ \citenamefont {Scalettar}}]{randeria_pairing_1992}%
  \BibitemOpen
  \bibfield  {author} {\bibinfo {author} {\bibfnamefont {M.}~\bibnamefont {Randeria}}, \bibinfo {author} {\bibfnamefont {N.}~\bibnamefont {Trivedi}}, \bibinfo {author} {\bibfnamefont {A.}~\bibnamefont {Moreo}},\ and\ \bibinfo {author} {\bibfnamefont {R.~T.}\ \bibnamefont {Scalettar}},\ }\href {https://doi.org/10.1103/PhysRevLett.69.2001} {\bibfield  {journal} {\bibinfo  {journal} {Phys. Rev. Lett.}\ }\textbf {\bibinfo {volume} {69}},\ \bibinfo {pages} {2001} (\bibinfo {year} {1992})}\BibitemShut {NoStop}%
\bibitem [{\citenamefont {Trivedi}\ and\ \citenamefont {Randeria}(1995)}]{trivedi_deviations_1995}%
  \BibitemOpen
  \bibfield  {author} {\bibinfo {author} {\bibfnamefont {N.}~\bibnamefont {Trivedi}}\ and\ \bibinfo {author} {\bibfnamefont {M.}~\bibnamefont {Randeria}},\ }\href {https://doi.org/10.1103/PhysRevLett.75.312} {\bibfield  {journal} {\bibinfo  {journal} {Phys. Rev. Lett.}\ }\textbf {\bibinfo {volume} {75}},\ \bibinfo {pages} {312} (\bibinfo {year} {1995})}\BibitemShut {NoStop}%
\bibitem [{\citenamefont {Sandvik}(1998)}]{sandvik_stochastic_1998}%
  \BibitemOpen
  \bibfield  {author} {\bibinfo {author} {\bibfnamefont {A.~W.}\ \bibnamefont {Sandvik}},\ }\href {https://doi.org/10.1103/PhysRevB.57.10287} {\bibfield  {journal} {\bibinfo  {journal} {Phys. Rev. B}\ }\textbf {\bibinfo {volume} {57}},\ \bibinfo {pages} {10287} (\bibinfo {year} {1998})}\BibitemShut {NoStop}%
\bibitem [{\citenamefont {Beach}(2004)}]{beach_identifying_2004}%
  \BibitemOpen
  \bibfield  {author} {\bibinfo {author} {\bibfnamefont {K.~S.~D.}\ \bibnamefont {Beach}},\ }\href {http://arxiv.org/abs/cond-mat/0403055} {\bibfield  {journal} {\bibinfo  {journal} {arXiv:cond-mat/0403055}\ } (\bibinfo {year} {2004})}\BibitemShut {NoStop}%
\bibitem [{\citenamefont {Shao}\ \emph {et~al.}(2017)\citenamefont {Shao}, \citenamefont {Qin}, \citenamefont {Capponi}, \citenamefont {Chesi}, \citenamefont {Meng},\ and\ \citenamefont {Sandvik}}]{shao_nearly_2017}%
  \BibitemOpen
  \bibfield  {author} {\bibinfo {author} {\bibfnamefont {H.}~\bibnamefont {Shao}}, \bibinfo {author} {\bibfnamefont {Y.~Q.}\ \bibnamefont {Qin}}, \bibinfo {author} {\bibfnamefont {S.}~\bibnamefont {Capponi}}, \bibinfo {author} {\bibfnamefont {S.}~\bibnamefont {Chesi}}, \bibinfo {author} {\bibfnamefont {Z.~Y.}\ \bibnamefont {Meng}},\ and\ \bibinfo {author} {\bibfnamefont {A.~W.}\ \bibnamefont {Sandvik}},\ }\href {https://doi.org/10.1103/PhysRevX.7.041072} {\bibfield  {journal} {\bibinfo  {journal} {Phys. Rev. X}\ }\textbf {\bibinfo {volume} {7}},\ \bibinfo {pages} {041072} (\bibinfo {year} {2017})}\BibitemShut {NoStop}%
\bibitem [{\citenamefont {Shao}\ and\ \citenamefont {Sandvik}(2023)}]{shao_progress_2023}%
  \BibitemOpen
  \bibfield  {author} {\bibinfo {author} {\bibfnamefont {H.}~\bibnamefont {Shao}}\ and\ \bibinfo {author} {\bibfnamefont {A.~W.}\ \bibnamefont {Sandvik}},\ }\href {https://doi.org/10.1016/j.physrep.2022.11.002} {\bibfield  {journal} {\bibinfo  {journal} {Phys. Rep.}\ }\textbf {\bibinfo {volume} {1003}},\ \bibinfo {pages} {1} (\bibinfo {year} {2023})}\BibitemShut {NoStop}%
\bibitem [{\citenamefont {Jiang}\ \emph {et~al.}(2022)\citenamefont {Jiang}, \citenamefont {Liu}, \citenamefont {Klein}, \citenamefont {Wang}, \citenamefont {Sun}, \citenamefont {Chubukov},\ and\ \citenamefont {Meng}}]{WLJiang2022}%
  \BibitemOpen
  \bibfield  {author} {\bibinfo {author} {\bibfnamefont {W.}~\bibnamefont {Jiang}}, \bibinfo {author} {\bibfnamefont {Y.}~\bibnamefont {Liu}}, \bibinfo {author} {\bibfnamefont {A.}~\bibnamefont {Klein}}, \bibinfo {author} {\bibfnamefont {Y.}~\bibnamefont {Wang}}, \bibinfo {author} {\bibfnamefont {K.}~\bibnamefont {Sun}}, \bibinfo {author} {\bibfnamefont {A.~V.}\ \bibnamefont {Chubukov}},\ and\ \bibinfo {author} {\bibfnamefont {Z.~Y.}\ \bibnamefont {Meng}},\ }\href {https://doi.org/10.1038/s41467-022-30302-x} {\bibfield  {journal} {\bibinfo  {journal} {Nat. Commun.}\ }\textbf {\bibinfo {volume} {13}},\ \bibinfo {pages} {2655} (\bibinfo {year} {2022})}\BibitemShut {NoStop}%
\bibitem [{\citenamefont {Hao}\ \emph {et~al.}(2022)\citenamefont {Hao}, \citenamefont {Pan}, \citenamefont {Sun}, \citenamefont {Meng},\ and\ \citenamefont {Qi}}]{YCHao2022}%
  \BibitemOpen
  \bibfield  {author} {\bibinfo {author} {\bibfnamefont {Y.}~\bibnamefont {Hao}}, \bibinfo {author} {\bibfnamefont {G.}~\bibnamefont {Pan}}, \bibinfo {author} {\bibfnamefont {K.}~\bibnamefont {Sun}}, \bibinfo {author} {\bibfnamefont {Z.~Y.}\ \bibnamefont {Meng}},\ and\ \bibinfo {author} {\bibfnamefont {Y.}~\bibnamefont {Qi}},\ }\href {https://doi.org/10.1088/0256-307X/39/9/097102} {\bibfield  {journal} {\bibinfo  {journal} {Chin. Phys. Lett.}\ }\textbf {\bibinfo {volume} {39}},\ \bibinfo {pages} {097102} (\bibinfo {year} {2022})}\BibitemShut {NoStop}%
\bibitem [{\citenamefont {Challa}\ and\ \citenamefont {Landau}(1986)}]{challa_critical_1986}%
  \BibitemOpen
  \bibfield  {author} {\bibinfo {author} {\bibfnamefont {M.~S.~S.}\ \bibnamefont {Challa}}\ and\ \bibinfo {author} {\bibfnamefont {D.~P.}\ \bibnamefont {Landau}},\ }\href {https://doi.org/10.1103/PhysRevB.33.437} {\bibfield  {journal} {\bibinfo  {journal} {Phys. Rev. B}\ }\textbf {\bibinfo {volume} {33}},\ \bibinfo {pages} {437} (\bibinfo {year} {1986})}\BibitemShut {NoStop}%
\bibitem [{\citenamefont {Isakov}\ and\ \citenamefont {Moessner}(2003)}]{isakov_interplay_2003}%
  \BibitemOpen
  \bibfield  {author} {\bibinfo {author} {\bibfnamefont {S.~V.}\ \bibnamefont {Isakov}}\ and\ \bibinfo {author} {\bibfnamefont {R.}~\bibnamefont {Moessner}},\ }\href {https://doi.org/10.1103/PhysRevB.68.104409} {\bibfield  {journal} {\bibinfo  {journal} {Phys. Rev. B}\ }\textbf {\bibinfo {volume} {68}},\ \bibinfo {pages} {104409} (\bibinfo {year} {2003})}\BibitemShut {NoStop}%
\bibitem [{\citenamefont {Zuo}\ \emph {et~al.}(2021)\citenamefont {Zuo}, \citenamefont {Yin}, \citenamefont {Cao},\ and\ \citenamefont {Zhong}}]{zuo_scaling_2021}%
  \BibitemOpen
  \bibfield  {author} {\bibinfo {author} {\bibfnamefont {Z.}~\bibnamefont {Zuo}}, \bibinfo {author} {\bibfnamefont {S.}~\bibnamefont {Yin}}, \bibinfo {author} {\bibfnamefont {X.}~\bibnamefont {Cao}},\ and\ \bibinfo {author} {\bibfnamefont {F.}~\bibnamefont {Zhong}},\ }\href {https://doi.org/10.1103/PhysRevB.104.214108} {\bibfield  {journal} {\bibinfo  {journal} {Phys. Rev. B}\ }\textbf {\bibinfo {volume} {104}},\ \bibinfo {pages} {214108} (\bibinfo {year} {2021})}\BibitemShut {NoStop}%
\end{thebibliography}%

\end{document}